\documentclass[11pt]{article}
\usepackage[margin=1in]{geometry}
\usepackage[ruled, vlined, linesnumbered]{algorithm2e}
\usepackage{amsmath}
\usepackage{amsfonts}
\usepackage{amsthm}
\usepackage{bbm}
\usepackage[colorlinks = true, citecolor = blue, urlcolor = blue]{hyperref}
\usepackage[font=footnotesize,labelfont=bf]{caption}
\usepackage{enumitem}
\usepackage{blindtext}
\usepackage{titlesec}
\usepackage{titletoc}
\usepackage{todonotes}
\usepackage{graphicx}
\usepackage[square]{natbib}
\usepackage{cleveref}
\usepackage{subcaption}
\usepackage{siunitx}
\usepackage{authblk}

\title{A Fast and Scalable Pathwise-Solver for Group Lasso and Elastic Net Penalized Regression via Block-Coordinate Descent}
\author[1]{James Yang}
\author[1]{Trevor Hastie}
\affil[1]{Department of Statistics, Stanford University}
\date{\today}

\newcommand{\sA}{\mathcal{A}}

\newcommand{\sS}{\mathcal{S}}

\newcommand{\pr}[1]{\left(#1\right)}
\newcommand{\br}[1]{\left[#1\right]}

\newcommand{\R}{\mathbb{R}}

\newcommand{\norm}[1]{\left\|#1\right\|}
\newcommand{\abs}[1]{\left|#1\right|}
\newcommand{\set}[1]{\left\{#1\right\}}

\newcommand{\Normal}{\mathcal{N}}

\newcommand{\zeros}{\mathbf{0}}
\newcommand{\ones}{\mathbf{1}}

\newcommand{\NA}{\text{NA}}
\DeclareMathOperator{\diag}{diag}
\DeclareMathOperator{\vecop}{vec}

\DeclareMathOperator{\subjto}{subject\;to}
\DeclareMathOperator*{\minimize}{minimize\;}

\DeclareMathOperator*{\argmin}{argmin}

\newtheorem{theorem}{Theorem}[section]

\begin{document}
\maketitle

\abstract{%
We develop fast and scalable algorithms based on block-coordinate descent
to solve the group lasso and the group elastic net
for generalized linear models along a regularization path.
Special attention is given when the loss is the usual least squares loss (Gaussian loss).
We show that each block-coordinate update can be solved efficiently
using Newton's method and further improved using an adaptive bisection method,
solving these updates with a quadratic convergence rate.
Our benchmarks show that our package \texttt{adelie} performs 3 to 10 times faster
than the next fastest package on a wide array of both simulated and real datasets.
Moreover, we demonstrate that our package is a competitive lasso solver as well,
matching the performance of the popular lasso package \texttt{glmnet}.
}

\section{Introduction}\label{sec:introduction}

The group lasso \citep{yuan:2006} is an extension of 
the celebrated lasso \citep{Tibshirani1996}
that aims to additionally capture group structure within the features to induce group-wise sparsity in the solutions.
Specifically, the group lasso tends to zero-out \emph{groups of coefficients} rather than individual coefficients like the lasso.
The group lasso is a bona fide extension in that if all groups are of size one,
then we recover the lasso.
Despite the wide popularity of the lasso for its interpretability of its sparse solutions,
there is a drawback when multiple columns of a feature matrix represent one entity.
For example, a categorical feature, or a factor, is often represented
as indicators for each category, yielding many columns.
In such cases, the group lasso will select all or none of these columns
whereas the lasso may select any subset of them,
leaving the user in an ambiguous situation to decide whether the factor
is informative or not \citep{meier:2008}.
Moreover, the result of lasso changes based on the encoding scheme of factors, 
which is an undesirable property \citep{yuan:2006}.
For these reasons, the group lasso has gained popularity over the years.

While the group lasso seems very close in spirit to the lasso
(as both models induce sparsity in their solutions),
their computational aspects are vastly different.
A popular method for solving the lasso is coordinate-descent \citep{friedman:2010}.
The authors demonstrate that coordinate-descent can be made extremely efficient
with a closed-form solution to each coordinate update.
As we shall see in \Cref{sec:algorithm}, the group lasso does not enjoy
such properties as soon as a group size is greater than one.
Still, inspired by the success of coordinate-descent with the lasso, 
many authors pursue (block) coordinate-descent as the primary workhorse of their algorithm
\citep{yuan:2006,meier:2008,tseng:2001,sparsegl:2022}.
To solve each (block) coordinate update,
people have gravitated towards proximal gradient methods such as
ISTA, FISTA, and FISTA with adaptive-restarts 
\citep{sparsegl:2022,beck:2009,klosa:2020,wright:2009,loris:2009,sls:2016,odonoghue:2015}.
These methods are attractive for problems like the group lasso
because the block-coordinate updates can be solved iteratively
where each iteration has a simple closed-form solution,
now reminiscient of the lasso.

In this paper, we develop fast and scalable algorithms for solving the group lasso
under the usual least squares loss (Gaussian loss)
as well as any twice-continuously differentiable convex, proper, and closed losses.
In particular, we can solve generalized linear models 
such as logistic regression, poisson regression, and multinomial regression
with the group lasso penalty.
Following the work of \citet{zou:2005}, which extends the lasso to the elastic net,
we work with an analogous elastic net penalty for the group lasso.
Our work largely follows the work of \citet{friedman:2010}
and our main result lies in the block-coordinate update.
Specifically, we propose a novel algorithm to solve each block-coordinate update
via Newton's method, which can be shown to have a convergence guarantee and a quadratic convergence rate.
We provide a publicly available Python package \texttt{adelie}
with brief installation directions in \Cref{sec:discussion}.

In \Cref{sec:algorithm}, we present our main result for solving
the group elastic net under the Gaussian loss using block-coordinate descent
with a special emphasis on the block-coordinate update.
\Cref{sec:glm} extends this result to general convex losses
using proximal quasi-Newton method.
\Cref{sec:multi} demonstrates that we may then use our results to easily 
fit multi-response data as well with no special changes to the solver.
Finally, \Cref{sec:benchmark} shows our benchmark results comparing
our package to existing R packages for solving the group lasso 
as well as the lasso.
\section{Preliminaries and Notations}\label{sec:notations}

Let $\R$, $\R^n$, and $\R^{m\times n}$ denote the space of real numbers,
real vectors of size $n$, and real matrices of shape $m\times n$, respectively.
Similarly, we define the analogous spaces for the non-negative real numbers $\R_+$
and the positive real numbers $\R_{++}$.
We denote $\ell_p$-norms on $\R^n$ as 
$\norm{x}_p := (\sum_{i=1}^n \abs{x_i}^p)^{1/p}$.
We write $\norm{x}_W := \sqrt{x^\top W x}$ to be the $\ell_2$-norm
scaled by weights $W \in \R_+^{n\times n}$.
We denote $\zeros$ and $\ones$ to be the vector of zeros and ones, respectively.
The identity matrix is denoted as $I_n \in \R^{n \times n}$ 
and $I$ if the dimensions are implicitly defined based on context.
We define $[n] := \set{1,\ldots, n}$ to be the set of integers 
ranging from $1$ to $n$.
For any vector $v \in \R^n$ and a set of indices $S \subseteq [n]$,
let $v_S$ be the subset of $v$ along indices in $S$.
If $S$ is a subset of $G$ groups, that is, $S \subseteq[G]$
and each group $g \in [G]$ has size $p_g$ such that $\sum_{g=1}^G p_g = p$,
then $v_S$ is the subset of $v$ along each group so that
$v_S = \set{v_g \in \R^{p_g} : g \in S}$.
In either context, we denote $v_{-S} \equiv v_{S^c}$ as the subset of $v$
along indices (groups) not in $S$.
For a real-valued function $f : \R^n \to \R$,
we denote its gradient as $\nabla f$ and the hessian as $\nabla^2 f$.
For two square matrices $A, B \in \R^{n\times n}$,
we denote $A \preceq B$ if $B-A$ is positive semi-definite
and $A \prec B$ if $B-A$ is positive definite.
We denote $A \otimes B$ to be the Kronecker product of any two arbitrary matrices $A$ and $B$.
\section{Algorithms for Group Lasso/Elastic Net under Gaussian Loss}\label{sec:algorithm}

Consider the usual regression setup where 
$X \in \R^{n\times p}$ and $y\in \R^n$ are the feature matrix and response vector, respectively.
Let $G$ be the number of groups and $p_g$ be the size of group $g \in [G]$
so that $X$ may be decomposed as
$\br{X_1 \, X_2 \, \ldots \, X_G}$ where $X_g \in \R^{n \times p_g}$ is the feature matrix
corresponding to group $g$ features only.
In this section, we discuss our main algorithm 
to solve the group elastic net problem under the Gaussian loss given by
\begin{align}
    \minimize_{(\beta_0,\beta) \in \R^{1+p}} \quad&
    \frac{1}{2} \norm{y - X \beta - \beta_0 \ones}_W^2 + \lambda P_{\alpha, \omega}(\beta)
    \label{eq:algorithms:grpnet-gaussian}
\end{align}
and 
\begin{align}
    P_{\alpha, \omega}(\beta)
    :=
    \sum\limits_{g=1}^G \omega_g \pr{
        \alpha \norm{\beta_g}_2 + \frac{1-\alpha}{2} \norm{\beta_g}_2^2
    }
    \label{eq:algorithms:grpnet-penalty}
\end{align}
where $\beta_g \in \R^{p_g}$ are the coefficients corresponding to group $g$.
Here, $W \in \R_+^{n\times n}$ is assumed to be a diagonal matrix where
the diagonal specifies the observation weights.
Moreover, we assume that the sum of the diagonal $\ones^\top W \ones$ is exactly $1$
so that the weights are normalized.
Typically, we use the uniform weights, which results in $W \equiv n^{-1} I$.
The penalty factor $\omega \in \R_+^G$ allows us to
control the \emph{relative} penalty across the groups and $\lambda \geq 0$
to control the \emph{overall} penalty.
A common strategy is to set $\omega_g \equiv \sqrt{p_g}$ so that
the groups are penalized relative to their size \citep{yuan:2006}. 
The elastic net parameter $\alpha \in [0,1]$ dictates the proportion
of the group lasso penalty ($\norm{\cdot}_2$) and the ridge penalty ($\norm{\cdot}_2^2$)
so that $\alpha$ closer to $1$ induces more sparsity in the solutions.
Note that the intercept $\beta_0$ is unpenalized.
Hence, without loss of generality, we assume for simplicity 
that $X$ is column-centered and $y$ is centered 
(where the center is a weighted average using the diagonal of $W$ as weights)
so that we may replace $\beta_0 \equiv 0$.
This simplification is specific to the Gaussian loss and
we will soon remove this assumption when we move to general 
convex losses in \Cref{sec:glm}.

Although the groups typically have less correlation \emph{between} groups than \emph{within} groups,
as most applications tend to group correlated features together,
it is nonetheless still possible to have strong correlation between groups.
In this case, the group lasso can still suffer from the same erratic behavior observed with the lasso
where the group lasso will tend to pick one of the correlated groups and ignore the rest.
For this reason, the group elastic net penalty can help ameliorate
this extreme behavior by applying some ridge penalty
to shrink every group towards each other. 

\begin{figure}[t]
    \centering 
    \includegraphics[width=\linewidth]{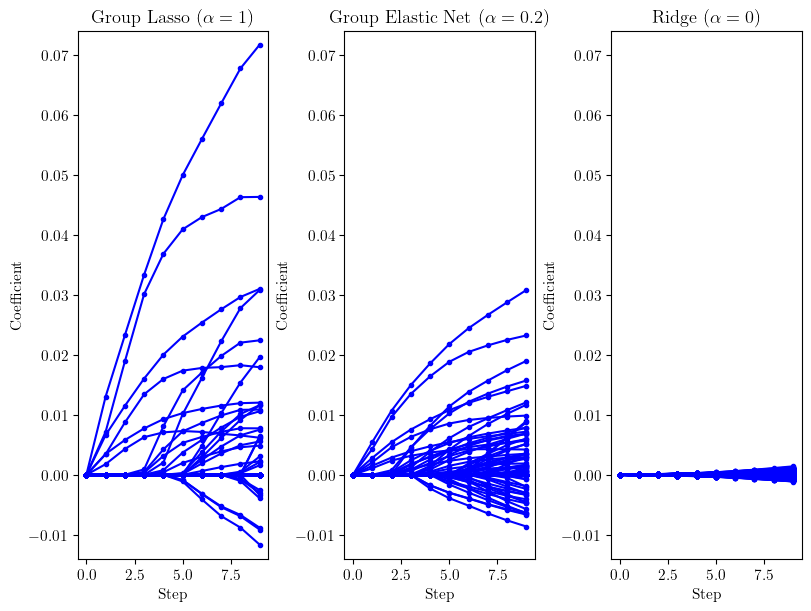}
    \caption{A comparison of the coefficient profile for the group lasso, group elastic net,
    and the ridge on the Leukemia dataset by \citep{Golub1999}.
    From left to right, we gradually see more features entering the model
    with an overall shrinkage on the coefficients towards zero
    until the ridge includes every feature.
    }
    \label{fig:algorithm:leukemia}
\end{figure}

Following \citep{friedman:2010},
we show in \Cref{fig:algorithm:leukemia} a comparison of the solution paths
for the group lasso, group elastic net, and ridge on the Leukemia dataset from \citet{Golub1999}\footnote{https://www.openintro.org/data/index.php?data=golub}.
This dataset contains 72 samples and 7129 gene expression probes that are measured with DNA microarrays.
To introduce group structure, we take each column of gene probes $X_{\cdot j}$
and create two more features $X_{\cdot,j}^2$, $X_{\cdot,j}^3$,
that is, the square and cube of the original column (element-wise).
We then group every three such features together.
The response $y$ is a binary variable indicating cancer or not for each observation.
However, for the sake of illustration, we treat it as a continuous variable.
We run the group lasso ($\alpha=1$), group elastic net ($\alpha=0.2$),
and ridge ($\alpha=0$) for $100$ values of $\lambda$, and plot
the coefficient profiles for the first $10$ values.
The total time to run all three models for all $100$ values of $\lambda$ was $0.5$ seconds
on a standard M1 Macbook Pro.
It is clear that group lasso tends to pick sparse models with few large coefficients
whereas the ridge includes all the features but shrinks them towards zero.
The group elastic net is somewhere in between where we observe more non-zero coefficients
than the group lasso but fewer than the ridge.

Our method to solve~\labelcref{eq:algorithms:grpnet-gaussian} is based on
the block-coordinate descent (BCD) algorithm~\citep{tseng:2001}.
Since the objective is composed of a smooth, convex, closed, and proper component 
and a separable non-differentiable component,
BCD provably attains the minimizer.
The BCD algorithm cycles through each group $g=1,\ldots, G$
and solves~\labelcref{eq:algorithms:grpnet-gaussian} only for $\beta_g$
while fixing all other coefficients.
In other words, we solve the following optimization problem:
\begin{align}
    \minimize_{\beta_g\in \R^{p_g}} \quad&
    \frac{1}{2} \beta_g^\top X_g^\top W X_g \beta_g
    - (y - \sum_{g' \neq g} X_{g'} \beta_{g'})^\top W X_g \beta_g
    + \lambda \omega_g \pr{\alpha \norm{\beta_g}_2 + \frac{1-\alpha}{2} \norm{\beta_g}_2^2}
    \label{eq:algorithm:bcd}
\end{align}
As we show in \Cref{ssec:algorithm:newton}, 
it is advantageous to first perform an eigen-decomposition of $X_g^\top W X_g = Q_g \Lambda_g Q_g^\top$
and transform $\beta_g \mapsto Q_g^\top \beta_g$.
Here, $Q_g \in \R^{p_g \times p_g}$ is a square orthogonal matrix
and $\Lambda_g \in \R_+^{p_g \times p_g}$ is a non-negative diagonal matrix,
which may contain some zeros.
Using the fact that the penalty is invariant under rotation
and letting $\tilde{\beta}_g := Q_g^\top \beta_g$,
\labelcref{eq:algorithm:bcd} changes to
\begin{align}
    \minimize_{\tilde{\beta}_g\in \R^{p_g}} \quad&
    \frac{1}{2} \tilde{\beta}_g^\top \Lambda_g \tilde{\beta}_g
    - (y - \sum_{g' \neq g} X_{g'} \beta_{g'})^\top W X_g Q_g \tilde{\beta}_g
    + \lambda \omega_g \pr{\alpha \norm{\tilde{\beta}_g}_2 + \frac{1-\alpha}{2} \norm{\tilde{\beta}_g}_2^2}
    \label{eq:algorithm:bcd-transform}
\end{align}
We abstract this problem into the following general problem
\begin{align}
    \minimize_{x \in \R^{d}} \quad&
    \frac{1}{2} x^\top \Sigma x - v^\top x
    + \lambda \norm{x}_2
    \label{eq:algorithm:bcd-general}
\end{align}
where $\Sigma \in \R_+^{d \times d}$ is a non-negative diagonal matrix
and $\lambda \geq 0$.
Relating back to~\labelcref{eq:algorithm:bcd-transform},
we have $\Lambda_g + \lambda \omega_g (1-\alpha) I \mapsto \Sigma$,
$Q_g^\top X_g^\top W (y - \sum_{g'\neq g} X_{g'} \beta_{g'}) \mapsto v$,
and $\lambda \omega_g \alpha \mapsto \lambda$ (with a slight abuse of notation).
The rest of the section focuses on solving~\labelcref{eq:algorithm:bcd-general}.
Since $\lambda > 0$ is the only interesting case,
we assume without loss of generality that $\lambda > 0$.

We make a few remarks on~\labelcref{eq:algorithm:bcd-general}.
Suppose for the moment that the observation weights $W = n^{-1} I$ are equal weights
to be consistent with the other references.
In the literature, we have found that most, if not all, methods
only consider isotropic $\Sigma$, that is, $\Sigma = t I$ for some scalar $t > 0$.
\citet{yuan:2006} directly assume that 
$X_g$ in~\labelcref{eq:algorithm:bcd} has orthonormal columns
so that $\Sigma \equiv I$. 
In the general case, they at least assume
$X_g$ to be full-rank so that it admits a QR decomposition with an invertible $R$.
Then, they solve~\labelcref{eq:algorithm:bcd} with $X_g$ replaced with the $Q$ matrix
and naively rescale the result by $R^{-1}$.
This strategy has drawbacks because it assumes full-rankness
and, more fundamentally, the resulting coefficient is \emph{not} the solution to
the original problem~\labelcref{eq:algorithm:bcd}.
This was also observed and noted in~\citet{simon:2012}
and the authors proposed to fix the issue by amending the group lasso penalty
from $\norm{\beta_g}_2$ to $\norm{X_g \beta_g}_2$.
While this allows the derivation to work out nicely,
we otherwise view this change merely as a mathematical convenience
and not a satisfactory solution to the original problem~\labelcref{eq:algorithm:bcd}.
Other authors such as \citet{sparsegl:2022,beck:2009,meier:2008}
majorize $X_g^\top X_g$ using an isotropic matrix.
They are able to show convergence and a convergence rate of $O(1/k^2)$
where $k$ is the number of iterations.
However, we noticed in practice that their methods converge very slowly 
either when $X_g^\top X_g$ is near singular or the dimension of the block increases.
As the block-coordinate update is the most critical component of the BCD algorithm,
we desire a more reliable and a faster algorithm.
Our method directly solves~\labelcref{eq:algorithm:bcd-general}
with no workarounds or approximations and 
works in full generality for any non-negative diagonal $\Sigma$,
or equivalently, any \emph{arbitary} $X_g$.
Moreover, we attain a quadratic convergence rate and a stable algorithm in practice.

It was previously noted that we must perform an eigen-decomposition for $X_g^\top W X_g$.
This decomposition only needs to be done \emph{at most once per group}.
Moreover, we will see later in \Cref{ssec:algorithm:screen-active} that 
the decomposition only needs to be done on a small subset of the groups.

We conclude this section with the statement of \Cref{thm:algorithm:existence}
(see \Cref{app:thm:algorithm:existence} for the proof).
We state this theorem for two reasons.
First, the problem in~\labelcref{eq:algorithm:bcd-general} is indeed more general
than~\labelcref{eq:algorithm:bcd-transform}.
If $\Sigma$ is not positive definite, then for a specific choice of $v$,
\labelcref{eq:algorithm:bcd-general} may not have a solution due to an unbounded objective.
Since we will now restrict our attention to \labelcref{eq:algorithm:bcd-general},
we provide a sufficient condition for the existence of a solution for completion.
It is easy to check via the SVD of $X_g$
that the conditions of \Cref{thm:algorithm:existence}
hold for our original problem~\labelcref{eq:algorithm:bcd}.
Secondly, condition \labelcref{eq:algorithm:bcd-general-ass1}
will become crucial in \Cref{ssec:algorithm:newton-abs}
where we develop our adaptive bisection method.

\begin{theorem}[Sufficient Condition for the Existence of the Minimizer of \labelcref{eq:algorithm:bcd-general}]
\label{thm:algorithm:existence} 
Consider the optimization problem of~\labelcref{eq:algorithm:bcd-general}
with $\lambda > 0$.
If the following condition holds
\begin{align}
    v_S = \zeros
    ,\quad
    S := \set{i \in [d]: \Sigma_{ii} = 0}
    \label{eq:algorithm:bcd-general-ass1} 
\end{align}
then a finite minimizer exists.
Moreover, the minimizer is given by
\begin{align}
    x^\star
    &=
    \begin{cases}
    \pr{\Sigma + \frac{\lambda}{\norm{x^\star}_2} I}^{-1} v ,& \norm{v}_2 > \lambda \\
    0 ,& \norm{v}_2 \leq \lambda
    \end{cases}
    \label{eq:algorithm:block-solution}
\end{align}
\end{theorem}

\subsection{Vanilla Newton's Method-Based Algorithm}\label{ssec:algorithm:newton}

We now discuss how to solve \labelcref{eq:algorithm:newton:norm-solution} via Newton's method.
Assume \labelcref{eq:algorithm:bcd-general-ass1} holds so that a solution exists by \Cref{thm:algorithm:existence}.
Without loss of generality, 
we only consider the first case of \labelcref{eq:algorithm:block-solution} when $\norm{v}_2 > \lambda$.
Without knowing $\norm{x^\star}_2$, there is no closed-form solution for $x^\star$.
In an endeavor to find an expression for $\norm{x^\star}_2$,
we take the $\ell_2$-norm on both sides to get that $\norm{x^\star}_2$ must satisfy
\begin{align}
    \sum\limits_{i=1}^d
    \frac{v_i^2}{(\Sigma_{ii} \norm{x^\star}_2 + \lambda)^2}
    =
    1
    \label{eq:algorithm:newton:norm-solution}
\end{align}
Unfortunately, there is no closed-form solution for $\norm{x^\star}_2$ from \labelcref{eq:algorithm:newton:norm-solution} either.
We see this as the main difference from performing coordinate-descent for lasso.
Namely, the coordinate update in lasso has a simple closed-form solution (soft-thresholding function) \citep{friedman:2010}.
However, as soon as a group has size greater than 1, 
there is no closed-form solution to its block-coordinate update in the group lasso.
In spite of this difficulty, we shall see shortly that a simple Newton's method 
can numerically solve \labelcref{eq:algorithm:newton:norm-solution} efficiently with guaranteed convergence.
In turn, once we find $\norm{x^\star}_2$, we may substitute it
into \labelcref{eq:algorithm:block-solution} 
to get the full solution $x^\star$ in $O(d)$ time.
Note that we can achieve this complexity because $\Sigma$ is diagonal.
Indeed, the subsequent arguments will also heavily rely on this fact
to achieve $O(dk)$ time complexity, where $k$ is the number of Newton iterations,
which would otherwise be $\Omega(d^2 k)$ if $\Sigma$ were not diagonal.
This computation reduction was the primary reason for transforming
our problem from \labelcref{eq:algorithm:bcd}
to~\labelcref{eq:algorithm:bcd-transform}.
We also note in passing that
\citet[Ex. 4.7]{sls:2016} also discovered the relationship between
\labelcref{%
eq:algorithm:block-solution,%
eq:algorithm:newton:norm-solution%
} from which
they proposed to solve for $\norm{x^\star}_2$ via the golden-section search.

Now, define $\varphi : [0, \infty) \to \R$ by
\begin{align}
    \varphi(h)
    &:=
    \sum\limits_{i=1}^d
    \frac{v_i^2}{(\Sigma_{ii} h + \lambda)^2}
    - 1
    \label{eq:algorithm:newton:varphi-def}
\end{align}
so that \labelcref{eq:algorithm:newton:norm-solution} is now 
a one-dimensional root-finding problem for $\varphi$.
In~\Cref{app:algorithm:newton:varphi-def},
we show that $\varphi$ is strictly decreasing, strictly convex,
and $\varphi(0) > 0 > \lim_{h\to\infty} \varphi(h)$,
so that a root (uniquely) exists.
Our strategy is then to apply Newton's method to
search for the root of $\varphi$.
Newton's method starts with an initial point $h^{(0)}$
and iteratively updates
\begin{align}
    h^{(k+1)} = h^{(k)} - \varphi'(h^{(k)})^{-1} \varphi(h^{(k)})
    \label{eq:algorithm:newton:newton}
\end{align}
\Cref{alg:algorithm:newton:newton} summarizes Newton's method.
By \Cref{thm:algorithm:newton:newton-convex},
Newton's method has a guaranteed convergence to the root of $\varphi$
for any initial point $h^{(0)} \geq 0$ such that $\varphi(h^{(0)}) \geq 0$
(e.g. $h^{(0)} \equiv 0$).
For a proof of \Cref{thm:algorithm:newton:newton-convex},
see \Cref{app:thm:algorithm:newton:newton-convex}.
Moreover, since $\varphi$ is strongly convex with Lipschitz second derivative,
we also have a quadratic convergence rate~\citep{boyd:2004}.

\begin{theorem}[Convergence Result of Newton's Method]%
\label{thm:algorithm:newton:newton-convex}
Suppose $f : \R \mapsto \R$ is convex, 
continuously differentiable with a non-vanishing derivative, and decreasing.
Suppose further that there exists a root $r \in \R$ such that $f(r) = 0$.
Let $x^{(0)}$ be any initial point such that $f(x^{(0)}) \geq 0$.
Then, the Newton iterates $\set{x^{(k)}}_{k=0}^\infty$ defined by
\begin{align}
    x^{(k+1)} = x^{(k)} - f'(x^{(k)})^{-1} f(x^{(k)})
    \label{eq:thm:algorithm:newton:newton}
\end{align}
is an increasing sequence that converges to a root of $f$.
\end{theorem}

\begin{algorithm}[t]
    \caption{Newton's Method}\label{alg:algorithm:newton:newton}
    \KwData{$\Sigma$, $v$, $\lambda$, $\varepsilon$, $h^{(0)}$}
    $h \gets h^{(0)}$\;
    \While{$\abs{\varphi(h)} > \varepsilon$} {
        $h \gets h - \frac{\varphi(h)}{\varphi'(h)}$\;
    }
    return $h$\;
\end{algorithm}

It is clear from the discussion so far why our method, in principle, 
is faster and more stable than the aforementioned proximal gradient methods
such as ISTA, FISTA, and FISTA with adaptive restarts.
First, none of these methods are able to give any 
convergence rate comparable to our quadratic convergence rate
largely because they are first-order methods.
Secondly, while these proximal gradient methods search the entire $\R^d$ space,
we have reduced our problem to a \emph{one-dimensional} root finding problem.
Lastly, we have established a well-behaving sequence of iterates 
(e.g. increasing and bounded).

\begin{figure}[t]
    \centering
    \includegraphics[width=\linewidth]{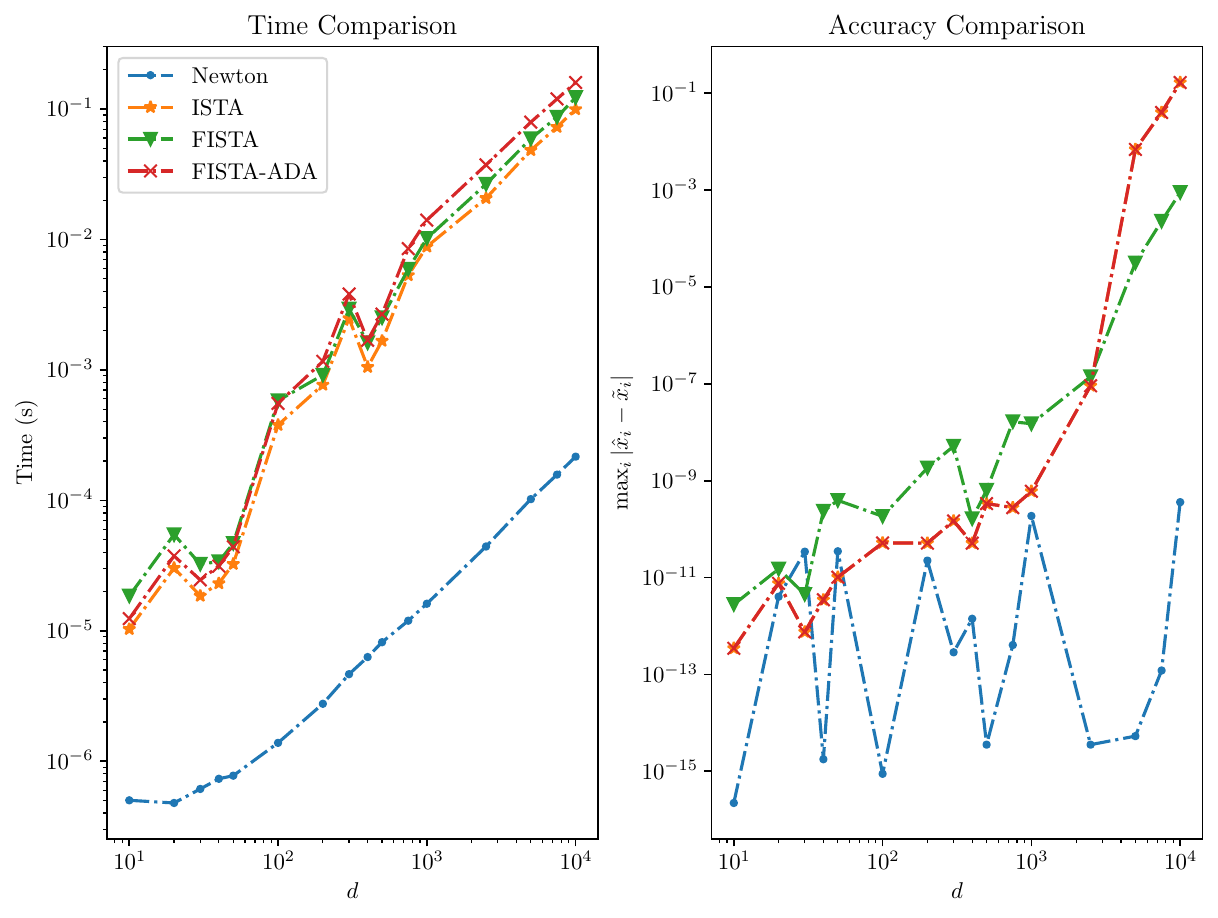}
    \caption{Time and accuracy comparison between the proximal gradient methods 
    and our Newton's method.
    Overall, the proximal gradient methods suffer from large computational overhead due to large number of iterations,
    and fails to converge as the dimension increases.
    Contrastingly, Newton's method is quite stable, properly converges, and runs 10 to 1000 times faster.
    }
    \label{fig:algorithm:newton:pgd-newton}
\end{figure}

\Cref{fig:algorithm:newton:pgd-newton} 
shows a comparison of the proximal gradient methods and our Newton method,
clearly demonstrating the superiority of our method.
We implemented all these methods as optimized as possible in C++.
For FISTA and FISTA with adaptive restarts, 
we follow the details laid out by the authors of
\citet{beck:2009,odonoghue:2015}.
We generate the diagonal of $\Sigma \in \R_{++}^{d \times d}$
from the uniform distribution on the interval $(0,1)$,
the linear term $v = \Sigma^{1/2} v'$ where $v' \sim \Normal(0, I_d)$,
and $\lambda = 0.1$.
We measure accuracy as
\begin{align}
    &\max\limits_{i=1,\ldots,d} 
    \abs{\hat{x}_i - \tilde{x}_i}
    \label{eq:algorithm:newton:accuracy}
\end{align}
where $\hat{x}$ is the solution returned by a solver
and $\tilde{x}$ is defined by~\labelcref{eq:algorithm:block-solution}
where the $\norm{x}_2$ is replaced with $\norm{\hat{x}}_2$.
Note that the proximal methods fail to converge within 10000 iterations as $d$ increases,
and even when they converge, Newton's method performs 10 to 1000 times faster.
Furthermore, the accuracy of the Newton's method remains steady regardless of the dimension $d$,
while the proximal methods have trouble searching the space efficiently,
suffering from the lack of guidance from the curvature information.

\begin{figure}[t]
    \centering 
    \includegraphics[width=\linewidth]{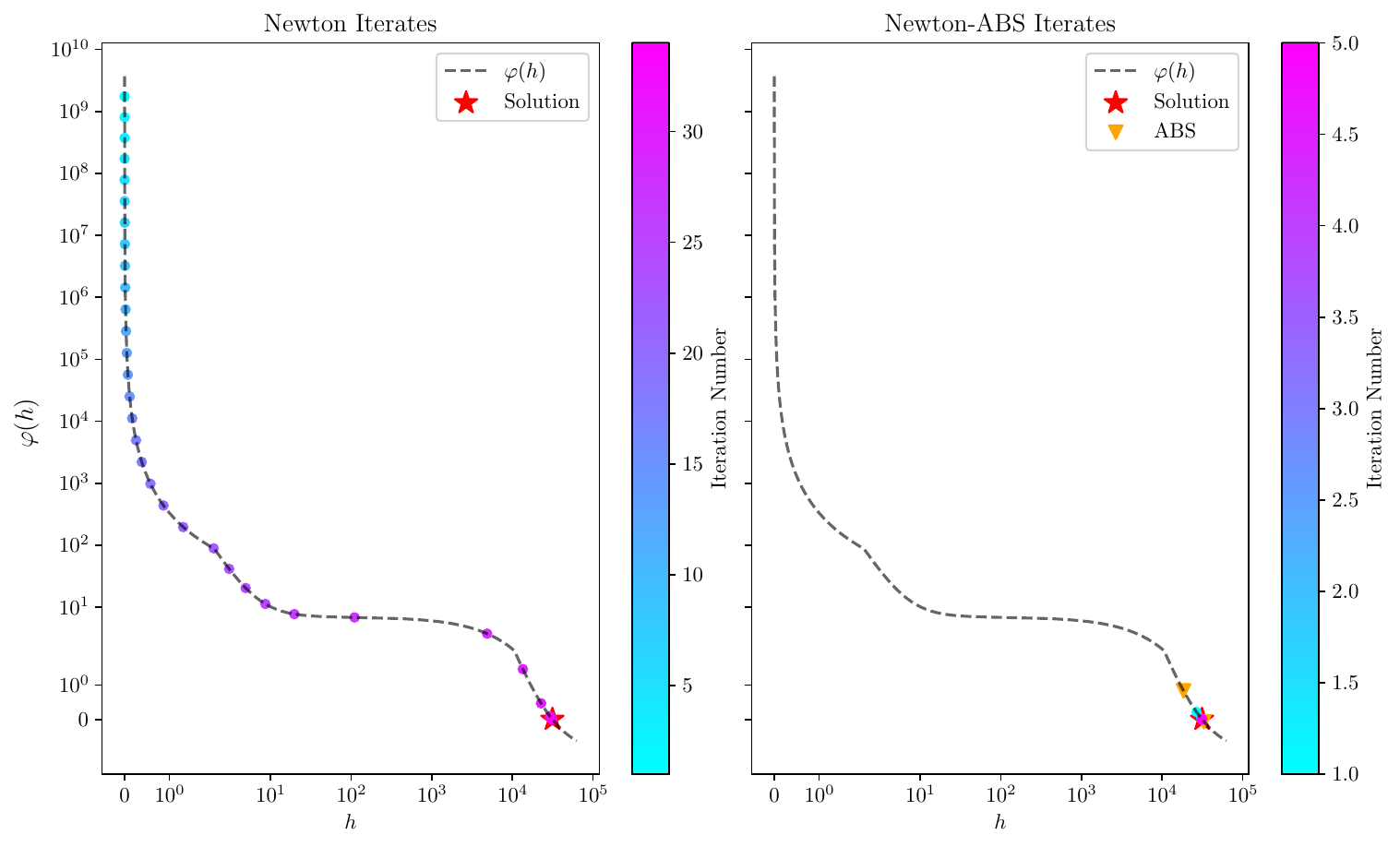}
    \caption{Plot of the iterates for the vanilla Newton's method (left) and Newton-ABS (right).
        In both plots, we display the iterates by their value of $\varphi$
        and describe their path by the color gradient scheme.
        The iterates are guaranteed to move from left to right.
        The solution is marked with a star.
        For visualization purposes, the axes were scaled using the symmetrical log transformation,
        which deceptively makes $\varphi$ look non-convex.
        It is clear that Newton's method struggles 
        where there is a sharp decay near the origin as the Newton iterates
        slowly exit the kink.
        Newton-ABS gets around this problem by finding a good initial point
        sufficiently away from the origin
        using our adaptive bisection strategy.
    }
    \label{fig:algorithm:newton:stuck}
\end{figure}

Despite the success of our Newton's method approach,
there is still room for improvement.
It is widely known that descent methods (including Newton's method)
is highly sensitive to the initialization.
A poor initialization may lead to significantly slower convergence in practice
while an initialization close to the root will be rewarded with incredible speed.
It then remains to show how to properly choose the initial starting point $h^{(0)}$.
First, it is always advantageous to select $h^{(0)}$ such that $\varphi(h^{(0)}) \geq 0$
to preserve the increasing property of the iterates and the convergence guarantees.
We have chosen $h^{(0)} \equiv 0$ thus far,
but it may not be the best initial value
as we may observe relatively larger number of iterations,
especially if $\Sigma$ is singular and $\lambda$ is small.
The left plot in \Cref{fig:algorithm:newton:stuck} shows an example of such a case.
We see that $\varphi$ decays rapidly near $h \approx 0$ 
and has an extremely flat tail.
The iterates slowly move from the origin until $h \approx 100$
and the speed picks up from there.
One possible remedy is to simply start Newton's method from a large value of $h$ rather than $0$.
However, as \Cref{fig:algorithm:newton:stuck} suggests,
because the function usually has a very flat tail,
Newton's method will likely shoot the iterate to the left with a large magnitude.
As a result, we may end up with an iterate that is negative.
Upon projecting back to $[0,\infty)$, we return to the situation
where we start Newton's method at the origin.
For these reasons, we develop in~\Cref{ssec:algorithm:newton-abs}
a special modification of our Newton-based approach, 
which describes our most performant and robust 
method that reduces iterations in nearly all configurations of the inputs.

\subsection{Newton's Method with Adaptive Bisection Starts Algorithm}\label{ssec:algorithm:newton-abs}

In this section, we improve our vanilla Newton's method-based algorithm described in~\Cref{ssec:algorithm:newton}.
The key idea is to first apply a bisection method to search for a good initial point.
In particular, based on \Cref{fig:algorithm:newton:stuck},
it is important to avoid the region of fast decay when $h$ is close to $0$,
as this is the region with many Newton iterations.
Once we find a point sufficiently close to the root,
we apply Newton's method as a refinement procedure, 
recovering all the convergence guarantees.
Although one could use any bisection method in theory,
we propose an \emph{adaptive bisection method} that has been most effective in our experience.
We note in passing that we have experimented with other bisection methods
such as the naive one that bisects evenly
and other sophisticated methods such as Dekker's method and Brent's method
\citep{brent:2013,dekker:1969}.
Brent's method, in particular, was the only competitive method to our approach,
however the runtime was quite unpredictable as it was too dependent
on the input values.

We first discuss how to find lower and upper bounds $h_\star$, $h^\star \in [0,\infty)$,
respectively, such that the root lies in $[h_\star, h^\star]$.
To motivate the derivation,
note that the root-finding problem for $\varphi$
is equivalent to finding the largest value $h > 0$ such that $\varphi(h) \geq 0$, or
\begin{align}
    \sup\set{
        h > 0 :
        \varphi(h) \geq 0
    }
    \label{eq:newton-abs:lower-bound-problem}
\end{align}
The idea is to find a different function $\varphi_{\star}(\cdot)$ such that 
\begin{align*}
    h_\star 
    :=
    \sup\set{
        h > 0:
        \varphi_{\star}(h) \geq 0
    }
    \implies
    \varphi(h_\star) \geq 0
\end{align*}
and $h_\star$ is easily computable.
Similarly, we find a function $\varphi^\star(\cdot)$ such that
\begin{align*}
    h^\star
    :=
    \inf\set{
        h > 0:
        \varphi^\star(h) \leq 0
    }
    \implies
    \varphi(h^\star) \leq 0
\end{align*}
This shows that the root must lie in $[h_\star, h^\star]$.
In \Cref{app:algorithm:newton-abs:lower-upper},
we find suitable functions $\varphi_\star$ and $\varphi^\star$,
and derive the following expressions for $h_\star$ and $h^\star$:
\begin{align}
    h_\star 
    &=
    \pr{
        \frac{
            -\lambda \ones^\top \Sigma \ones
            + \sqrt{(\ones^\top \Sigma \ones)^2 - (\ones^\top \Sigma^2 \ones) (\lambda^2 d - \norm{v}_1^2)}
        }{\ones^\top \Sigma \ones}
    }_+
    \label{eq:algorithm:newton-abs:lower}
    \\
    h^\star
    &= 
    \sqrt{
        \sum\limits_{i: \Sigma_{ii} > 0} v_i^2 \Sigma_{ii}^{-2}
    }
    \label{eq:algorithm:newton-abs:upper}
\end{align}

We now describe our proposed adaptive bisection method.
Ideally, we would like to know if the root is closer to $h_\star$ or $h^\star$.
If we believe that the root is much closer to $h_\star$,
we do not have to bisect $[h_\star, h^\star]$ at the mid-point, 
but perhaps at a point closer to $h_\star$.
Likewise, if the root were much closer to $h^\star$, 
we would like to bisect closer to $h^\star$.
Since we do not know the root a priori, it is not possible 
to determine which side it is closer to.
However, we can make an informed guess based on a \emph{prior} of how likely
the root is closer to $h_\star$, for example.
With this in mind, we construct our prior 
based on the derivation of $h^\star$ in \Cref{app:algorithm:newton-abs:lower-upper}.
For completeness, we provide the relevant part of the proof:
\begin{align}
    \sum\limits_{i=1}^d
    \frac{v_i^2}{(\Sigma_{ii} h + \lambda)^2}
    &=
    \sum\limits_{i: \Sigma_{ii} > 0}
    \frac{v_i^2}{(\Sigma_{ii} h + \lambda)^2}
    \leq 
    h^{-2}
    \sum\limits_{i: \Sigma_{ii} > 0}
    \frac{v_i^2}{\Sigma_{ii}^2 }
    \label{eq:algorithm:newton-abs:upper-approx}
\end{align}
It is easy to see that our $h^\star$ makes the right-side of 
\labelcref{eq:algorithm:newton-abs:upper-approx} equal to 1.
We note that the inequality above becomes tighter as $h^\star$ approaches the root.
This motivates us to consider the worst approximation error (rate)
\begin{align*}
    w
    := 
    \max\limits_{i: \Sigma_{ii} > 0} \frac{\lambda}{\Sigma_{ii} h^\star + \lambda}
    =
    \frac{\lambda}{\Sigma_\star h^\star + \lambda}
    \in 
    (0,1)
    ,\quad
    \Sigma_\star := \min\limits_{i : \Sigma_{ii} > 0} \Sigma_{ii}
\end{align*}
If $w \approx 0$, the inequality in~\labelcref{eq:algorithm:newton-abs:upper-approx} is tight,
which suggests that the root is close to $h^\star$.
Hence, $w$ represents the prior that the root is close to $h_\star$.
We bisect at the new point $h := wh_\star + (1-w)h^\star$.
If $\varphi(h) > \epsilon$ (sufficiently positive), we use $h^{(0)} := h$ as the initial point for our Newton's method.
Otherwise if $\varphi(h) < -\epsilon$ (sufficiently negative), we set $h^\star := h$ and repeat the bisection argument.
Finally, if $\abs{\varphi(h)} \leq \epsilon$ (sufficiently close to 0),
we declare $h$ to be the root.
\Cref{alg:algorithm:newton-abs:abs,alg:algorithm:newton-abs:nabs} summarize this procedure,
which we call \emph{Newton's Method with Adaptive Bisection Starts} (Newton-ABS).
We emphasize that we still have a convergence guarantee
since we eventually apply Newton's method.

\begin{algorithm}[t]
    \caption{Adaptive Bisection}\label{alg:algorithm:newton-abs:abs}
    \KwData{$\Sigma$, $v$, $\lambda$, $\varepsilon$}
    Compute $h_\star$ using \labelcref{eq:algorithm:newton-abs:lower}
    and $h^\star$ using \labelcref{eq:algorithm:newton-abs:upper}\;
    $\Sigma_\star \gets \min\limits_{i : \Sigma_{ii} > 0} \Sigma_{ii}$\;
    $h \gets h^\star$\;
    \While{$\varphi(h) < -\varepsilon$}{
        $h^\star \gets h$\; 
        $w \gets \frac{\lambda}{\Sigma_\star h^\star + \lambda}$\;
        $h \gets wh_\star + (1-w)h^\star$\;
    }
    return $h$;
\end{algorithm}

\begin{algorithm}[t]
    \caption{Newton's Method with Adaptive Bisection Starts (Newton-ABS)}\label{alg:algorithm:newton-abs:nabs}
    \KwData{$\Sigma$, $v$, $\lambda$, $\varepsilon$}
    $h \gets$ result of \Cref{alg:algorithm:newton-abs:abs}\;
    \If{$\abs{\varphi(h)} > \varepsilon$} {
        $h \gets$ result of \Cref{alg:algorithm:newton:newton} starting at $h^{(0)} = h$\;
    }
    $x^\star \gets (\Sigma + \lambda h^{-1} I)^{-1} v$\;
    return $x^\star$\;
\end{algorithm}

\Cref{alg:algorithm:newton-abs:nabs} can be further optimized.
If $h^\star - h_\star$ is below some threshold (e.g. $0.1$),
then we may skip bisection entirely and start Newton's method at $h^{(0)} = h_\star$.
This is because the adaptive bisection may move too slowly if the range is too small.
One may also enforce enough movement towards $h_\star$
by taking the max of $w$ with a minimal probability (e.g. $0.05$).
This will ensure that at least some proportion of $h_\star$ is taken 
if the prior suggests too strongly that the root is close to $h^\star$.
The idea is that it is better to overshoot towards $h_\star$ such that $\varphi$ is non-negative
so that we can apply Newton's method sooner.

\Cref{fig:algorithm:newton:stuck} shows a plot of the Newton iterations for 
the vanilla Newton's method from~\Cref{ssec:algorithm:newton} and our proposed Newton-ABS method.
It is clear that Newton's method struggles 
where there is a sharp decay near the origin as the Newton iterates
slowly exit the kink.
However, Newton-ABS gets around this problem due to the fact that the adaptive bisection
first moves in the opposite direction, starting at the upper bound $h^\star$
and moving towards $h_\star$, until $\varphi$ is non-negative.
In this example, only 2 adaptive bisections were required and 5 subsequent Newton iterations,
which is a noticeable improvement from the 34 vanilla Newton iterations.
Empirically, we have never observed a case where the initial value remains at the kink
after performing the adaptive bisection.

\subsection{Naive and Covariance Updates}\label{ssec:algorithm:naive-cov}

As in the coordinate descent algorithm for lasso,
we have a ``naive'' and ``covariance'' method to update
the residual or gradient vector, respectively
\citep{friedman:2010}.
Denote $\beta$ as the current coefficient vector
and suppose in the BCD algorithm, we are at group $g$.

The ``naive'' method keeps track of the residual $r := y - X \beta$.
To solve the BCD update~\labelcref{eq:algorithm:bcd},
we must prepare the inputs to \labelcref{eq:algorithm:bcd-transform}.
Recall that we can cache the eigen-decomposition of $X_g^\top W X_g$ 
so that we have access to its eigenvalues $\Lambda_g$ and eigenvectors $Q_g$.
Then, we may compute the linear term in the following way
\begin{align*}
    Q_g^\top X_g^\top W (y - \sum_{g' \neq g} X_{g'} \beta_{g'})
    =
    Q_g^\top X_g^\top W r + \Lambda_g Q_g^\top \beta_g
\end{align*}
In other words, we simply need to compute $\gamma_g := X_g^\top W r$,
express $\gamma_g$ and $\beta_g$ in the eigen-basis $Q_g$, 
and perform simple elementwise operations.
Altogether, the cost is $O((n+p_g)p_g)$.
Once the BCD update, $\tilde{\beta}_g$, is found by solving~\labelcref{eq:algorithm:bcd},
we must update the residual to reflect the change in $\beta_g$.
Given the current residual $r$, 
the change in residual, $\Delta r$, can then be expressed as
\begin{align*}
    \Delta r = -X_g \pr{\tilde{\beta}_g - \beta_g}
\end{align*}
So, the residual update requires another $O(n p_g)$ operations.
Sometimes, we can avoid performing the residual update entirely if we observe
that $\tilde{\beta}_g \equiv \beta_g$, i.e. the BCD update did nothing.
Nevertheless, the ``naive'' method has a worst-case cost of $O((n+p_g+k)p_g)$ per BCD update
where $k$ is the number of iterations for the Newton-ABS method.

The ``covariance'' method keeps track of the gradient $\gamma := X^\top W (y-X\beta)$.
This simplifies the input preparation for the BCD update~\labelcref{eq:algorithm:bcd},
since the linear term does not require the $O(np_g)$ operations to compute $X_g^\top W r$
as in the ``naive'' method so that the complexity is $O(p_g^2)$.
However, it could potentially make the gradient update more costly.
The change in the gradient $\Delta \gamma$ can be expressed as
\begin{align*}
    \Delta \gamma
    =
    - X^\top W X_g (\tilde{\beta}_g - \beta_g)
\end{align*}
where we only have access to $X^\top W X_g$ as a whole.
Hence, the gradient update requires $O(p\cdot p_g)$ operations.
Therefore, the ``covariance'' method has a worst-case cost of $O((p+p_g+k) p_g)$,
where $k$ is the number of iterations for the Newton-ABS method.

Depending on the use-case, the user may find that one method is faster than the other.
Specifically, when $n \ll p$, the ``naive'' method is more efficient,
and when $p \ll n$, the ``covariance'' method is more efficient.
However, in terms of memory, the ``naive'' method is almost always preferred
(unless $n \gg p \cdot n_{\sA}$ where $n_{\sA}$ is the number of active coefficients).
This is because the ``covariance'' method requires saving $X^\top W X_g \in \R^{p \times p_g}$
at the very least for the active groups $g$ (with non-zero coefficients).
With large $p$, this may be unmanageable.
While we could technically recompute $X^\top W X_g$ on-the-fly
if we have insufficient memory,
this would significantly hurt the performance and will fail to be competitive against the ``naive'' method.
On the other hand, the ``naive'' method only requires $O(n)$ extra memory for the residual.

\subsection{Matrix Abstraction}\label{ssec:algorithm:matrix}

The reader may notice that the BCD algorithm only ever interacts
with the matrix $X$ through a few operations,
namely, inner-products with columns of $X$.
Our software \texttt{adelie} takes advantage of this fact and abstracts out the matrix class.
Any matrix-like class supporting a few methods to compute such inner-products
can be passed into our solver.

This abstraction has been proven to be quite useful for many applications.
As shown in~\Cref{ssec:algorithm:naive-cov}, 
the bulk of the computation in each BCD update
lies in computing inner-products with $X$.
Hence, any effort into speeding up these inner-products
will yield large performance gains for the solver.
So, if $X$ is structured in some sense, it is extremely helpful
to exploit its structure to simplify the inner-product computations.
A simple example is when $X$ is a sparse matrix.
A more motivating example is in genome-wide association studies (GWAS)
where the feature matrix of interest is the genotype matrix of SNP information
taking on values in $\set{0,1,2,\NA}$.
Both the sparsity of this matrix as well as the restricted range of values
allow us to efficiently store this matrix with far more structure than a dense or a sparse matrix.
There is a plethora of references in this area, so we only mention a few
\citep{Uffelmann2021,Witte2010,Mills2019,He2022,He2021,He2024,Qian2020,Li2020}.
These genotype matrices can be of daunting size.
The UK biobank dataset \citep{Sudlow2015}, for example, contains 500,000 samples
and about 1.7 million SNPs (features).
Moreover, along with this genotype matrix, we also have a small set of dense covariates
such as age, sex, and 10 principal components of the genotype matrix.
Without the matrix abstraction, one must then represent the full feature matrix $X$ as a dense matrix,
which would require nearly $6$ terabytes of floating point numbers,
a size too large to load in the RAM of any modern computers.
Indeed, there has been some work to solve lasso on such large-scale data
\citep{Qian2020,Li2020,Li2021,He2024,zeng2018biglasso},
all implementing their own solvers for the purpose of
supporting their own representation of the large matrices.
Using our method, we can represent, for example, a dense matrix for the covariates,
a structured matrix for the genotype matrix, and a column-wise ``concatenated'' matrix of the two.
The sole purpose of the concatenated matrix is simply to delegate
the inner-products to the correct sub-matrix,
which is an inexpensive operation. 
If one wishes to represent the matrix differently,
one only needs to implement a small matrix class 
without implementing a separate solver.

We provide a last example specifically for the ``covariance'' method.
The covariance matrix $X^\top W X$ 
is usually a dense matrix, however, there are cases when it
can be approximated as a block-diagonal matrix \citep{He2024}.
This structure was exploited to greatly speed up the lasso solver
and save large amounts of memory, however, they implemented 
a separate lasso solver specifically assuming this structure.
The same optimization can be done with our software
with few lines of code just to represent the block-diagonal matrix.
The user may otherwise reuse our general solver.

\subsection{Pathwise Block-Coordinate Descent}\label{ssec:algorithm:pathwise}

We compute solutions along a decreasing sequence of the regularization parameter $\lambda$. 
We start with the smallest $\lambda$, say $\lambda_{\max}$, for which all 
groups with non-zero group lasso penalty have coefficients identically $\zeros$. 
This value can be directly computed once we fit all unpenalized variables. 
That is, if $U := \{g \in [G] : \omega_g \alpha = 0\}$ denotes the set of groups 
that have vanishing group lasso penalty, 
we first solve \labelcref{eq:algorithms:grpnet-gaussian} 
only for groups in $U$ to get their corresponding coefficients $\hat{\beta}_U$.
Then, we compute $\lambda_{\max}$ given by
\begin{align}
    \lambda_{\max}
    :=
    \max_{g : g \notin U}
    \frac{\norm{X_g^\top W (y - X_U \beta_U)}_2}{\alpha \omega_g}
\end{align}
By the KKT conditions, we have our desired property that the zero solution on groups not in $U$ is optimal at $\lambda_{\max}$.
Note that if $U$ is empty (which occurs if $\alpha = 0$ or every $\omega_g = 0$) so that every group is included in the model, then we may arbitrarily take any value for $\lambda_{\max}$. 
From $\lambda_{\max}$, we generate a decreasing sequence of $K$ number of $\lambda$ values 
down to $\epsilon \lambda_{\max}$ evenly-spaced on the log scale, 
where $\epsilon \in [0,1]$ is a user-specified value that determines the depth of the path. 
Following \citet{friedman:2010}, we typically set $\epsilon \in [0.0001, 0.01]$ and $K = 100$.

The advantage of a pathwise solver is that the solver is warm-started 
at each $\lambda$ (including $\lambda_{\max}$). 
This helps tremendously for smaller values of $\lambda$ 
where a cold-start generally makes the solver wander 
longer before reaching the optimum. 
This was also observed by \citet{friedman:2010}. 
Similar ideas exist in the convex literature, in particular, 
with interior-point methods and barrier methods \citep{boyd:2004}.

\subsection{Screen and Active Sets}\label{ssec:algorithm:screen-active}

In the BCD algorithm, we can layer two types of optimizations that 
both reduce computation by cycling over only a subset of the groups. 
The first optimization is based on a \emph{screening rule}, 
which tries to guess the set of groups that will likely contain 
the true active set (groups with non-zero coefficients).
We call such a set the \emph{screen set} and denote it as $\sS$. 
It is possible to modify the BCD algorithm 
to only ever iterate over the screen set.
Therefore, as mentioned in \Cref{sec:algorithm},
we may only cache the eigen-decomposition of $X_g^\top W X_g$ for $g \in \sS$.
However, this optimization comes at a (small) price of checking the KKT conditions.
The second optimization further reduces most of the computation 
to a subset of active groups in the screen set. 

In the first optimization, the screen rule can be a heuristic or a guarantee. 
The literature contains many types of such rules including
SAFE rule, strong rule, and EDPP rule
\citep{ghaoui2011safe,Tibshirani2012Strong,wang:2015}.
In particular, the SAFE rule and the EDPP rule are examples of guarantees
in the sense that at a fixed $\lambda$ if these rules discard groups,
then they guarantee that these groups will truly have zero coefficients.
On the other hand, the strong rule is an example of a heuristic, which does not
always guarantee that the groups that it discards will be inactive.
Despite the lack of a guarantee, we find that among these rules
the strong rule was the most effective one in practice
as it includes much fewer groups and is almost always correct.
For this reason, we implement the strong rule as part of our software.

We briefly describe the strong rule for the group elastic net problem. 
Suppose we have a solution $\tilde{\beta}$ at $\tilde{\lambda}$ and the current screen set is $\tilde{\sS}$. 
We must now solve for $\lambda < \tilde{\lambda}$ to obtain the next solution $\beta$. 
For the group elastic net, the strong rule searches 
for all remaining groups $g \notin \tilde{\sS}$ such that
\begin{align}
    \norm{X_g^\top W (y - X \tilde{\beta})}_2 < 2 \lambda - \tilde{\lambda}
    \label{eq:algorithm:screen-active:strong}
\end{align}
If this condition is satisfied, group $g$ is discarded from the screen set. 
Note that $2\lambda - \tilde{\lambda} = \lambda - (\tilde{\lambda} - \lambda)$
so that \labelcref{eq:algorithm:screen-active:strong} looks at the target $\lambda$
and a band below equal to the path gap at $\lambda$.
Once all such groups are discarded to obtain the next screen set $\sS$, 
we proceed to solve the group elastic net only over the groups in $\sS$,
effectively solving the (constrained) group elastic net problem
with the constraint that $\beta_{-\sS} = \zeros$;
at the start of the path, the screen set only contains unpenalized groups
so that the screen set is exactly the active set at $\lambda_{\max}$.
It is easy to imagine that the more accurate the screening rule is,
the smaller the screen set $\sS$, and therefore, the less computation
needed in the BCD algorithm.
Since we guess that groups outside the screen set are inactive, 
we must verify the KKT conditions to confirm our guess. 
This amounts to checking that
\begin{align*}
    \norm{X_g^\top W (y - X\beta)}_2 \leq \lambda
    ,\quad \forall g\notin \sS
\end{align*}
If the condition holds, then we may declare $\beta$ as the solution at $\lambda$.

The second optimization reduces the computation further by
restricting most of the BCD iterations to the current active set within the screen set.
Namely, we first iterate over the screen set once to determine the active groups.
Afterwards, if we have not converged, we iterate only over the active groups until convergence.
We then repeat the entire process until we reach convergence on the screen set.
This active-set strategy is also mentioned in 
\citet{friedman:2010,meier:2008,Krishnapuram2005}.

\subsection{Convergence Criterion}\label{ssec:algorithm:convergence}

In the BCD algorithm, we measure convergence by measuring
the average amount in which each block-update results in a change
in the linear prediction and take the largest such value.
Concretely, suppose $\tilde{\beta}$ is the current coefficient vector
and we have just updated group $g$ so that the new coefficient
$\beta$ only differs from $\tilde{\beta}$ in the entries defined by group $g$.
Then, we compute $p_g^{-1} \|X_g (\beta_g-\tilde{\beta}_g)\|_W^2$.
Note that since we have already cached $X_g^\top W X_g = Q_g \Lambda_g Q_g^\top$,
it is equivalent to computing $p_g^{-1} \|Q_g^\top (\beta_g - \tilde{\beta}_g)\|_{\Lambda_g}^2$.
Since we must apply the rotation $Q_g^\top$ 
on $\tilde{\beta}_g$ and $\beta_g$ anyways to solve \labelcref{eq:algorithm:bcd-transform},
computing the convergence measure is an easy task.
After one full cycle of the groups, we check if
\begin{align}
    \max\limits_{g}\quad
    p_g^{-1} \norm{X_g (\beta_g - \tilde{\beta}_g)}_W^2
    \leq
    \epsilon
    \label{eq:algorithm:convergence:metric}
\end{align}
where $\epsilon \geq 0$ is a user-specified tolerance level.
If the condition holds, we declare convergence.

\subsection{Lasso Optimization}\label{ssec:algorithm:lasso}

As mentioned before, groups of size $1$
have the same coordinate update as in the lasso,
which admits a closed-form solution.
Hence, at every coordinate update,
our software checks for the group sizes
and falls back to the lasso update in this special case.
Otherwise, we use our Newton-ABS method to solve the update.
\section{Algorithms for Regularized Generalized Linear Models}\label{sec:glm}

In \Cref{sec:algorithm}, we discussed our main algorithm
for solving the group elastic net under the Gaussian loss.
In this section, we extend our algorithm for general convex losses.

Suppose $\ell(\cdot) : \R^n \to \R$ is a convex, proper, and closed function
that is twice-continuously differentiable.
We typically consider $\ell(\cdot)$ as the negative log-likelihood of some GLM
so that is of the form
\begin{align*}
    \ell(\eta)
    :=
    \sum\limits_{i=1}^n
    w_i \pr{-y_i \eta_i + A(\eta_i)}
\end{align*}
where $A(\cdot)$ is the log-partition function of an exponential family.
However, for the current discussion, we need not restrict ourselves to such form.
Our goal is to solve the more general optimization problem
\begin{align}
    \minimize_{(\beta_0, \beta) \in \R^{1+p}} \quad&
    \ell(\eta)
    +
    \lambda P_{\alpha, \omega}(\beta)
    \label{eq:glm:grpnet}
    \\\subjto \quad&
    \eta = X\beta + \beta_0 \ones + \eta^0
    \nonumber
\end{align}
where the group elastic net penalty $P_{\alpha, \omega}(\cdot)$
is defined by \labelcref{eq:algorithms:grpnet-penalty}.
The only new term is $\eta^0 \in \R^n$, which we call the \emph{offset} vector.
The offset is a fixed vector that defines possibly different
baseline linear predictions for each observation so that
$X\beta + \beta_0 \ones$ is interpreted as the change in the linear prediction
from the given baseline.
For the Gaussian loss
\begin{align*}
    \ell(\eta)
    :=
    \sum\limits_{i=1}^n
    w_i \pr{-y_i \eta_i + \frac{\eta_i^2}{2}}
    \equiv
    \frac{1}{2} \pr{\norm{y - \eta}_W^2 - \norm{y}_W^2}
\end{align*}
the offset can be managed in a trivial manner by replacing $y$ with $y-\eta^0$
and running the algorithm described in \Cref{sec:algorithm} with no other changes.
However, with general loss functions, the offset can interact with the objective in a non-trivial manner.

We use proximal quasi-Newton method to solve \labelcref{eq:glm:grpnet}.
This method is also mentioned in \citet{friedman:2010,meier:2008}.
Moreover, \citet{Lee2014} proves a superlinear convergence rate
for proximal quasi-Newton methods (with line search).
However, as in \citet{friedman:2010}, we do not perform the line search
since it will significantly impact the overall runtime
and it does not seem necessary in practice thanks to the warm-starts 
in our pathwise solver.
We observe fast convergence, usually only requiring
1 to 5 iterations.

We are now in position to describe the proximal quasi-Newton method for our problem.
We begin with an initial value $(\beta_0^{(0)}, \beta^{(0)}) \in \R^{1+p}$.
For each iteration $k \geq 0$ until convergence,
we compute our current linear prediction
$\eta^{(k)} = X \beta^{(k)} + \beta_0^{(k)} \ones + \eta^0$.
We then perform a second-order expansion of $\ell(\cdot)$ around $\eta^{(k)}$
and solve for the proximal operator to get the next iterate
$(\beta_0^{(k+1)}, \beta^{(k+1)})$.
That is,
\begin{align}
    (\beta_0^{(k+1)}, \beta^{(k+1)})
    =
    \argmin_{(\beta_0, \beta) \in \R^{1+p}} \quad&
    \nabla \ell(\eta^{(k)})^\top \eta
    +
    \frac{1}{2} (\eta - \eta^{(k)})^\top W^{(k)} (\eta - \eta^{(k)})
    +
    \lambda P_{\alpha, \omega}(\beta)
    \label{eq:glm:prox}
    \\\quad\subjto\quad&
    \eta = X \beta + \beta_0 \ones + \eta^0
    \nonumber
\end{align}
Note that the linear term is indeed the gradient of $\ell(\cdot)$ at $\eta^{(k)}$;
however, $W^{(k)}$ need not be $\nabla^2 \ell(\eta^{(k)})$.
In general, our strategy is to find a positive diagonal matrix $W^{(k)}$
that majorizes $\nabla^2 \ell(\eta^{(k)})$, i.e. 
$0 \prec W^{(k)}$ diagonal and $\nabla^2 \ell(\eta^{(k)}) \preceq W^{(k)}$.
Of course, if $\nabla^2 \ell(\eta^{(k)})$ is already positive and diagonal,
we may simply take $W^{(k)} \equiv \nabla^2 \ell(\eta^{(k)})$.
The proximal operator \labelcref{eq:glm:prox} is simply
an instance of the group elastic net problem under the Gaussian loss
given by \labelcref{eq:algorithms:grpnet-gaussian}.
Hence, we may use our algorithm in \Cref{sec:algorithm} to solve \labelcref{eq:glm:prox}.
We summarize our algorithm in \Cref{alg:glm:prox}.
This algorithm is referred to as \emph{iteratively reweighted least squares} (IRLS)
in \citet{friedman:2010,McCullagh1989}.

\begin{algorithm}[t]
    \caption{Proximal Quasi-Newton Method for General Smooth, Convex Losses}\label{alg:glm:prox}
    \KwData{$X$, $y$, $\beta_0^{(0)}$, $\beta^{(0)}$, $\eta^0$}
    $k \gets 0$\;
    \While{not converged based on \labelcref{eq:glm:prox-convergence}}{
        $\eta^{(k)} = X\beta^{(k)} + \beta_0^{(k)} \ones + \eta^0$\;
        Compute $\nabla \ell(\eta^{(k)})$\;
        Compute $W^{(k)}$ appropriate for the given loss $\ell(\cdot)$\;
        Compute $(\beta_0^{(k+1)}, \beta^{(k+1)})$ by solving~\labelcref{eq:glm:prox}
        using our method in \Cref{sec:algorithm}\;
        $k \gets k + 1$\;
    }
    return $h$;
\end{algorithm}

\subsection{Application on Generalized Linear Models}\label{ssec:glm:glm}

Our primary application of \Cref{alg:glm:prox} is when $\ell(\cdot)$
represents the negative log-likelihood of a GLM.
Some examples include the Binomial family with the logit or probit link function, 
Poisson family with the log link function, and the Cox family.
In terms of the solver,
we do not make any special changes based on the loss function.
We note that this is one of the more notable deviations from the approach of \citet{friedman:2010},
as the authors made amends to the solver specialized for each GLM,
which we found unnecessary.

The only manner in which the loss function plays a role in~\Cref{alg:glm:prox}
is in computing the gradient $\nabla \ell(\eta^{(k)})$ and the hessian-like quantity $W^{(k)}$.
As a result, we have structured our software to abstract out
the loss function to a class that provides the gradient and the hessian,
similar to how we abstracted out the matrix classes (see \Cref{ssec:algorithm:matrix}).
As an example, for the Binomial family with the logit link function 
\citep{McCullagh1989} where the loss is given by 
\begin{align*}
    \ell(\eta)
    :=
    \sum\limits_{i=1}^n
    w_i \pr{-y_i \eta_i + \log(1 + \exp(\eta_i))}
\end{align*}
we implement a class to compute the gradient and hessian, respectively, given by
\begin{align}
    \nabla \ell(\eta)_i
    =
    -w_i (y_i - p(\eta_i))
    ,\quad
    \nabla^2 \ell(\eta)_{ij}
    =
    \begin{cases}
        w_i p(\eta_i) (1-p(\eta_i)) ,& i = j \\
        0 ,& i \neq j
    \end{cases}
    \label{eq:glm:binomial}
\end{align}
where $p(t) := 1 / (1 + \exp(-t))$ is the sigmoid function.

\subsection{Screen Sets and the KKT Check}\label{ssec:glm:screen}

We may reuse the strong rule mentioned in \Cref{ssec:algorithm:screen-active}
when solving \labelcref{eq:glm:grpnet}.
We follow the notation as in \Cref{ssec:algorithm:screen-active}.
The only change we make is in computing the scores
so that we discard group $g \notin \tilde{\sS}$ if
\begin{align*}
    \norm{X_g^\top \nabla \ell(X\tilde{\beta} + \tilde{\beta_0} \ones + \eta^0)}_2
    < 2 \lambda - \tilde{\lambda},
\end{align*}
Similarly, we check the KKT condition by verifying that
\begin{align*}
    \norm{X_g^\top \nabla \ell(X\beta + \beta_0 \ones + \eta^0)}_2
    \leq \lambda
    ,\quad \forall g \notin \sS
\end{align*}

\subsection{Numerical Stability Issues}\label{ssec:glm:numerical}

In \labelcref{eq:glm:prox}, it is possible that $W^{(k)}$ is nearly semi-definite,
which may lead to divergence issues.
For example, in the case of a Binomial family (logistic regression),
if the estimated probabilities $p_i^{(k)} := 1 / (1 + \exp(-\eta_i^{(k)}))$
are near $0$ or $1$, then $W_{ii}^{(k)}$ will be near $0$
(see \labelcref{eq:glm:binomial}).
Hence, we majorize $W^{(k)}$ one more time
by taking the maximum of each diagonal entry with a small positive number
(e.g. $10^{-12}$) to guarantee positiveness of $W^{(k)}$.
The authors of \citet{friedman:2010} only make this change for the Binomial family,
but we see no reason why this should not be done for any other loss functions.

\subsection{Convergence Criterion}\label{ssec:glm:convergence}

We discuss our convergence criterion.
At iteration $k \geq 0$, upon solving the proximal operator \labelcref{eq:glm:prox},
we check that
\begin{align}
    \abs{
        \pr{\eta^{(k+1)} - \eta^{(k)}}^\top
        \pr{\nabla \ell(\eta^{(k+1)}) - \nabla \ell(\eta^{(k)})}
    } 
    \leq 
    \epsilon \cdot n_{\sA}
    \label{eq:glm:prox-convergence}
\end{align}
and declare convergence if the condition holds.
Here, $n_{\sA}$ denotes the current number of coefficients (not groups) that were updated
and $\epsilon \geq 0$ is a user-specified tolerance level.
We explain our intuition behind this criterion below.

First, suppose we perform a first-order Taylor expansion of $\nabla \ell(\cdot)$ around $\eta^{(k)}$,
\begin{align*}
    \nabla \ell(\eta)
    \approx
    \nabla \ell(\eta^{(k)})
    +
    \nabla^2 \ell(\eta^{(k)}) 
    (\eta - \eta^{(k)})
\end{align*}
Then, we see that
\begin{align}
    \pr{\eta^{(k+1)} - \eta^{(k)}}^\top
    \pr{\nabla \ell(\eta^{(k+1)}) - \nabla \ell(\eta^{(k)})}
    \approx
    \pr{\eta^{(k+1)} - \eta^{(k)}}^\top
    \nabla^2 \ell(\eta^{(k)}) 
    \pr{\eta^{(k+1)} - \eta^{(k)}}
    \label{eq:glm:prox-convergence:quad}
\end{align}
which measures the change in linear predictions weighted accordingly 
based on the hessian at $\eta^{(k)}$.
This way, a change in the prediction along a direction
for which the hessian is small is downweighted, as this suggests
the loss $\ell(\cdot)$ is only mildly affected by such a change.
Conversely, a change along a direction for which the hessian is large
is upweighted, as this suggests the loss $\ell(\cdot)$ may be largely affected.

Next, we argue that \labelcref{eq:glm:prox-convergence}
is a generalization of the convergence metric in \Cref{ssec:algorithm:convergence}.
To see this, first replace $\eta = X\beta$, 
where we ignore the intercept and the offset momentarily for simplicity.
Then, from \labelcref{eq:glm:prox-convergence:quad} we have that
\begin{align}
    \pr{\eta^{(k+1)} - \eta^{(k)}}^\top
    \nabla^2 \ell(\eta^{(k)}) 
    \pr{\eta^{(k+1)} - \eta^{(k)}}
    =
    \Delta \beta^\top X^\top \nabla^2 \ell(\eta^{(k)}) X \Delta \beta
    \label{eq:glm:prox-convergence:quad-full}
\end{align}
where $\Delta \beta := \beta^{(k+1)} - \beta^{(k)}$.
Under the Gaussian loss where $\nabla^2 \ell(\eta^{(k)}) \equiv W$,
the right side of \labelcref{eq:glm:prox-convergence:quad-full} becomes
$\norm{X \Delta \beta}_W^2$.
Now, suppose (as in the BCD algorithm) that
$\beta^{(k+1)}$ only differs in one group from $\beta^{(k)}$.
That is, for some $g \in [G]$, $\beta^{(k+1)}_g \neq \beta^{(k)}_g$
and $\beta^{(k+1)}_{g'} \equiv \beta^{(k)}_{g'}$ for all $g' \neq g$.
Then, 
$\norm{X \Delta \beta}_W^2 = \norm{X_g \Delta \beta_g}_W^2$.
In \labelcref{eq:algorithm:convergence:metric},
we scale this quantity by $p_g$, which is
the total number of coefficients that were updated
(since we have fixed all other coordinates).
Hence, we scale by the analogous quantity in \labelcref{eq:glm:prox-convergence}.
\section{Algorithms for Multi-Response Data}\label{sec:multi}

In this section, we discuss methods to fit multi-response data. 
In multi-response data, we have a response \emph{matrix}
$y \in \R^{n \times c}$ rather than a response vector.
As a result, we change the group elastic net objective \labelcref{eq:glm:grpnet} to
\begin{align}
    \minimize_{\beta_0 \in \R^{c}, B \in \R^{p \times c}} \quad&
    \ell(\vecop(\eta^\top))
    + \lambda P_{\alpha, \omega}(\vecop(B^\top))
    \label{eq:multi:grpnet}
    \\\subjto\quad&
    \eta = X B + \ones \beta_0^\top + \eta^0
    \nonumber
\end{align}
where $\vecop(\cdot)$ is the operator that flattens a matrix into a one-dimensional array by stacking the columns.
Note that we must now fit a matrix of coefficients
along with an intercept vector $\beta_0 \in \R^c$ (one for each response).

While it is possible to specify any arbitrary group structure on $B$,
we focus only on two types: \emph{grouped} and \emph{ungrouped}.
The grouped method defines each \emph{row} of $B$ as a group
whereas the ungrouped method defines each \emph{coefficient} of $B$ as a group.
Hence, 
the grouped method yields a non-trivial group lasso problem
and the ungrouped method reduces to a lasso problem.
The penalty $P_{\alpha, \omega}$ simplifies to the following respective forms
\begin{align*}
    P_{\alpha, \omega}^{\text{grouped}}(\vecop(B^\top))
    &=
    \sum\limits_{g=1}^{p}
    \omega_g
    \pr{
        \alpha \norm{B_{g \cdot}}_2
        + \frac{1-\alpha}{2} \norm{B_{g \cdot}}_2^2
    }
    \\
    P_{\alpha, \omega}^{\text{ungrouped}}(\vecop(B^\top))
    &=
    \sum\limits_{g=1}^{p}
    \omega_g
    \pr{
        \alpha \norm{B_{g \cdot}}_1
        + \frac{1-\alpha}{2} \norm{B_{g \cdot}}_2^2
    }
\end{align*}

Although it may seem like \labelcref{eq:multi:grpnet} requires a different solver,
we now demonstrate that it can be easily solved with the tools built in \Cref{sec:algorithm,sec:glm}.
It is not difficult to show that the constraint in \labelcref{eq:multi:grpnet} can be rewritten as
\begin{align}
    \vecop(\eta^\top) = (X \otimes I_c) \vecop(B^\top) + (\ones \otimes I_c) \beta_0 + \vecop(\eta^{0\top})
\end{align}
Written this way, with 
the data matrix $[(\ones \otimes I_c) \quad (X \otimes I_c)] \in \R^{nc \times (1+p)c}$ and
the coefficient vector $(\beta_0, \vecop(B^\top)) \in \R^{(1+p)c}$,
we return to the setting of \labelcref{eq:glm:grpnet} (with no intercept).
As discussed in \Cref{ssec:algorithm:matrix},
as soon as we implement a matrix to represent the Kronecker product with an identity matrix,
we are able to solve \labelcref{eq:multi:grpnet} with our existing tools.
Such a matrix can be implemented to compute inner-products efficiently due to its structure.
For the Multigaussian family where the loss function is given by
\begin{align*}
    \ell(\eta)
    =
    \sum\limits_{i=1}^n
    w_i \pr{-\sum\limits_{j=1}^c y_{ij} \eta_{ij} + \frac{\norm{\eta_{i\cdot}}_2^2}{2}}
\end{align*}
we may simply rely on our algorithm that solves the group elastic net under the Gaussian loss (see \labelcref{eq:algorithms:grpnet-gaussian}).
Otherwise, we must rely on our algorithm that handles general convex losses (see \labelcref{eq:glm:grpnet}).

We make a special remark on the Multinomial family where the loss is given by
\begin{align*}
    \ell(\eta)
    :=
    \sum\limits_{i=1}^n
    w_i \pr{
        - \sum\limits_{j=1}^c y_{ij} \eta_{ij}
        + \log\pr{\sum\limits_{j=1}^c \exp(\eta_{ij})}
    }
\end{align*}
Unlike the GLMs discussed previously, this marks the first example
where the hessian $\nabla^2 \ell(\eta)$ is not exactly positive and diagonal.
Observe that the model is not identifiable in the sense that
for any fixed $\set{\beta_{0,j}, \beta_{\cdot, j}}_{j=1}^c$,
we have that $\set{\beta_{0,j}-d_0, \beta_{\cdot,j}-d}_{j=1}^c$ yields the same loss
for any $d_0 \in \R$ and $d \in \R^p$.
As a result, the hessian is never positive definite.
Moreover, the hessian is only block-diagonal.
Consequently, \citet{friedman:2010} approach the Multinomial family quite differently from the other GLMs.
First, they tweak the solver to cycle over the classes $j=1,\ldots, c$ while fixing all other class coefficients,
so that the gradient and hessian in \labelcref{eq:glm:prox}
is only with respect to the linear predictors for class $j$, namely, $\eta_{\cdot,j}$.
Because of this modification, given a candidate solution $(\beta_0, \beta)$,
they found it necessary to then further minimize only the lasso penalty 
with respect to $d_0$ and $d$ to improve the convergence speed.
Despite these efforts, we have found a real dataset on which
their software \texttt{glmnet} fails to converge.
We pursue a different and much simpler approach.
We simply need to construct a positive, diagonal matrix that majorizes $\nabla^2 \ell(\eta)$.
\citet{Simon2013ABD} show that $2 \diag(\nabla^2 \ell(\eta))$
is a valid choice for dominating $\nabla^2 \ell(\eta)$.
Hence, we set $W^{(k)} := 2 \diag(\nabla^2 \ell(\eta^{(k)}))$
in \labelcref{eq:glm:prox} and reuse our work from \Cref{sec:glm}.
With this approach, our software is able to converge.
This method implicitly settles the identifiability issue
precisely because $W^{(k)}$ is now positive definite
(or can be made positive definite as in \Cref{ssec:glm:numerical}).
Out of curiosity, we also experimented with minimizing the penalty
with respect to $d_0$ and $d$, 
but to our surprise, this resulted
in a less stable algorithm often diverging.
\section{Benchmark}\label{sec:benchmark}

In this section, we show an assortment of benchmarks
to demonstrate the power of our method.
We developed a highly efficient implementation of our algorithms discussed in
\Cref{sec:algorithm,sec:glm,sec:multi} in our package \texttt{adelie}
where most of the code is written in C++.
Although we intend \texttt{adelie} to be a Python-first package,
we are currently developing a R package as well\footnote{https://github.com/JamesYang007/adelie-r}.
We exported enough functionalities to perform the benchmark comparisons in this section.
We note that both the Python and R packages share the same C++ code,
which performs most of the computation,
and the only difference comes from the input preparation for the C++ solver.
Hence, there is only a negligible difference in their runtimes.
All benchmarks were run on an M1 Macbook Pro.

The packages we benchmark against are 
\texttt{sparsegl} \citep{sparsegl:2022,sparsegl:R:2023},
\texttt{gglasso} \citep{Yang2015-zx,gglasso:R:2024},
and \texttt{grpnet} \citep{Helwig:2024}.
We note that these packages are pathwise-solvers as well.
We attempted to include the following packages,
however, they either failed to converge or had substantially
longer runtimes:
\texttt{grplasso} \citep{meier:2008,grplasso:R:2020},
\texttt{SGL} \citep{Simon2013-yv,SGL:R:2019},
and \texttt{grpreg} \citep{Breheny2015-zj}.

For our real data analyses, we consider the following datasets:
\begin{itemize}
\item Leukemia dataset \citep{Golub1999}:
    the Leukemia dataset has been previously described in \Cref{sec:algorithm}.
    However, in this section, we will use it to classify whether an individual has Leukemia.
\item UJIIndoorLoc dataset~\citep{ujiindoorloc}:
    the UJIIndoorLoc dataset 
    contains WiFi fingerprints of Received Signal Strength Intensity (RSSI)
    from location trackers.
    We remove any redundant features from the dataset.
    Our task is to predict the longitude coordinate of the sample.
\item Prostate dataset \citep{Singh2002-gv}:
    the Prostate dataset 
    gene expressions from a microarray study of prostate cancer.
    Our task is to classify whether an individual has prostate cancer.
\end{itemize}
We summarize the datasets in \Cref{tab:benchmark:real}.

\begin{table}
\begin{center}
\begin{tabular}{c c c c}
    \hline
    Name & n & G & Task \\
    \hline
    Leukemia & 72 & 7129 & classification \\
    UJIIndoorLoc & 19937 & 465 & regression \\
    Prostate & 102 & 6033 & classification
\end{tabular}
\end{center}
\caption{%
The number of samples ($n$) and groups ($G$)
as well as the task for each of the real datasets.
}
\label{tab:benchmark:real}
\end{table}

\subsection{Group Lasso Comparisons}\label{ssec:benchmark:group-lasso}

\begin{figure}[t]
    \centering 
    \begin{subfigure}{0.95\textwidth}
        \includegraphics[width=\linewidth]{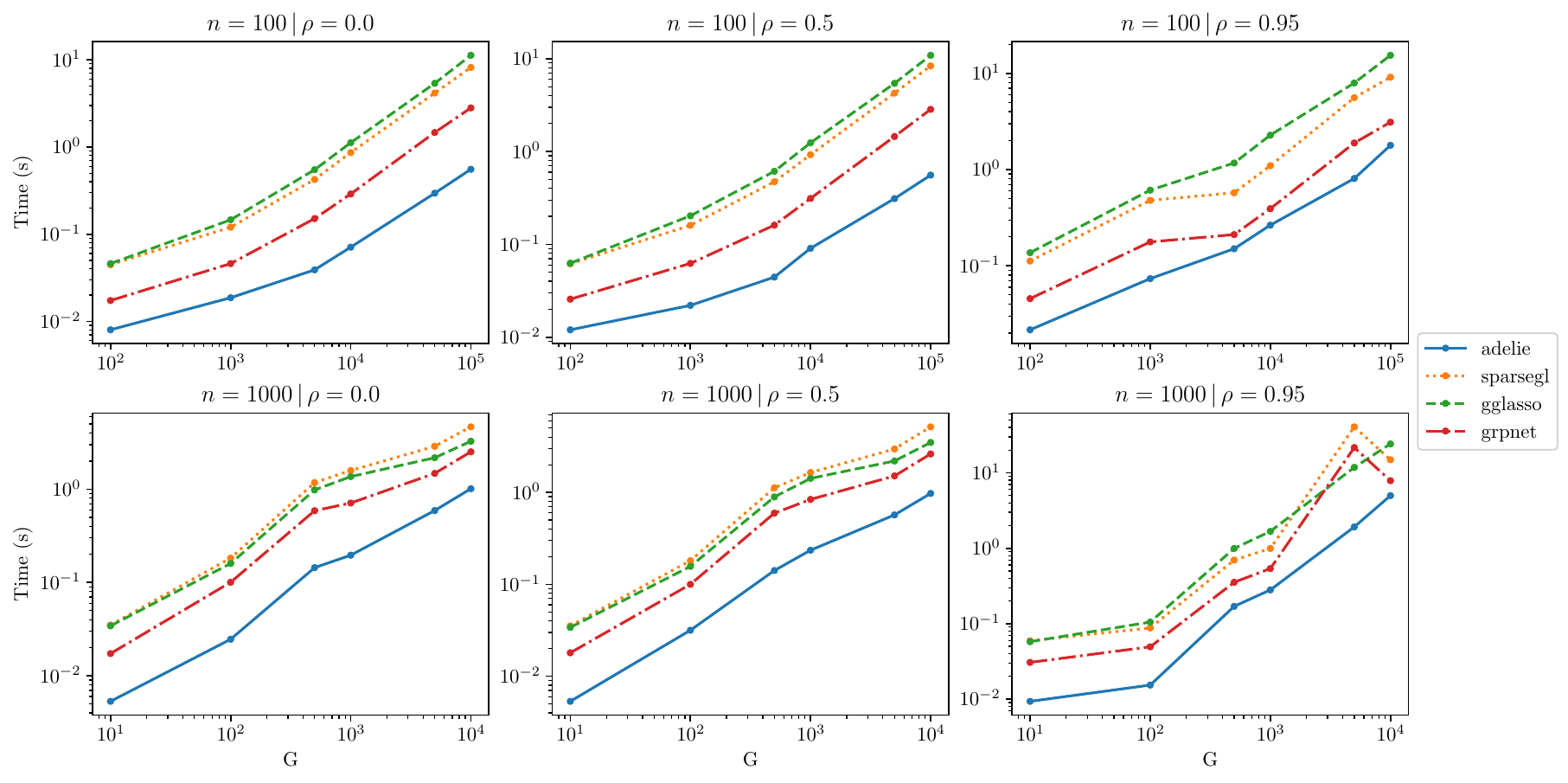}
        \caption{Gaussian Loss}
        \label{fig:benchmark:group-lasso-sim:gaussian}
    \end{subfigure}

    \begin{subfigure}{0.95\textwidth}
        \includegraphics[width=\linewidth]{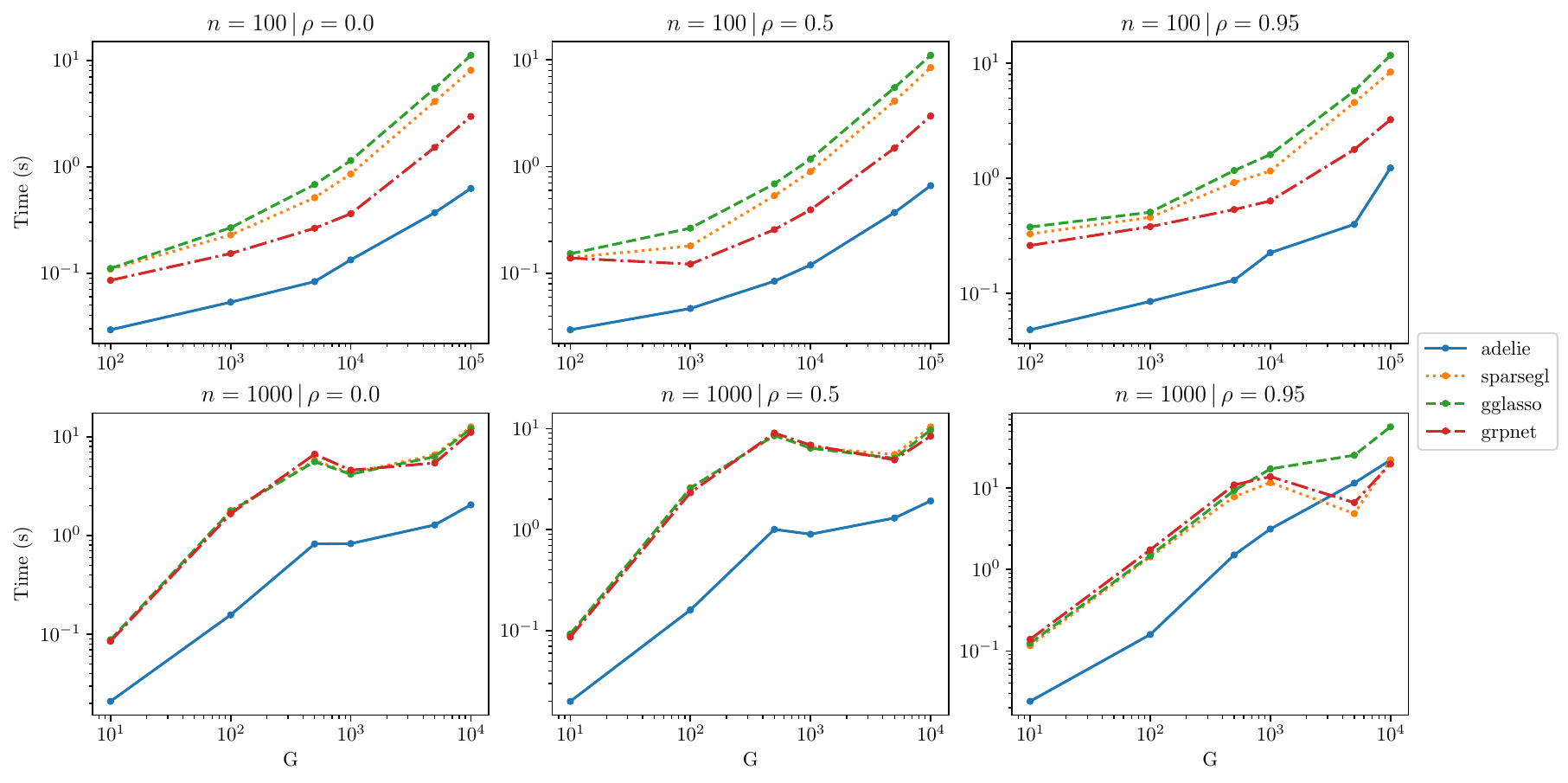}
        \caption{Binomial Loss}
        \label{fig:benchmark:group-lasso-sim:binomial}
    \end{subfigure}

    \caption{%
    Timing comparisons for solving the group lasso under 
    the Gaussian loss (\Cref{fig:benchmark:group-lasso-sim:gaussian})
    and the Binomial loss (\Cref{fig:benchmark:group-lasso-sim:binomial})
    against existing R packages.
    We study a small ($n=100$) and large ($n=1000$) sample size case,
    and for each case we vary the 
    (equi-)correlation of the features ($\rho$).
    }
    \label{fig:benchmark:group-lasso-sim}
\end{figure}

\begin{figure}[t]
    \centering 
    \includegraphics[width=\linewidth]{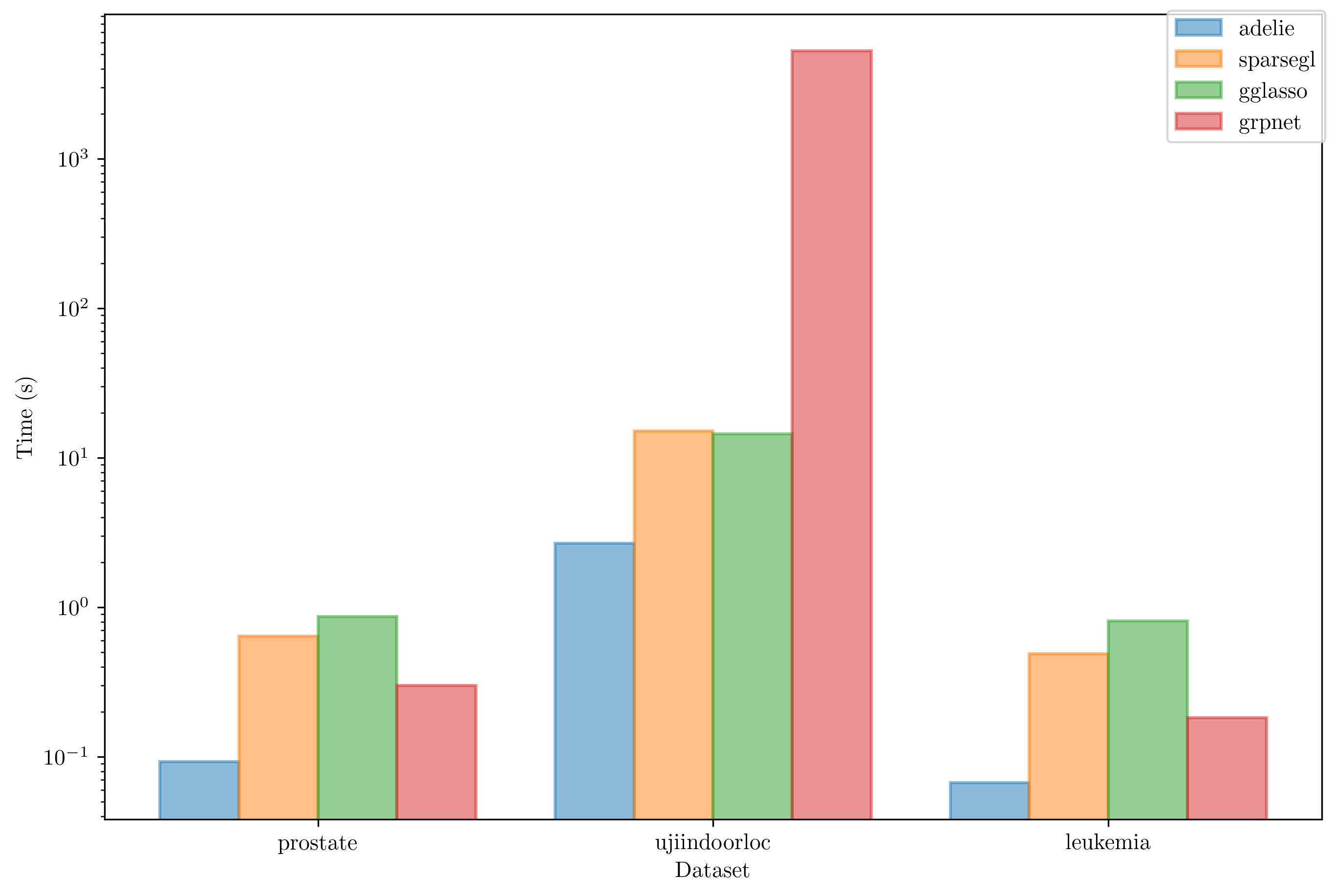}
    \caption{%
    Timing comparisons for solving the group lasso 
    for the real datasets against existing R packages.
    }
    \label{fig:benchmark:group-lasso-real}
\end{figure}

We show a comparison of solving the group lasso under the
Gaussian and Binomial losses using both simulated and real data.

We first describe our simulation setup.
Our setup is similar to the one used by \citet{yuan:2006}.
Let $n$ be the number of samples, $G$ be the number of groups,
and $\rho \in [0,1]$ be the equi-correlation between features.
We generate $Z_i \sim \Normal(0, I_G)$ for $i \in [n]$
and $W \sim \Normal(0, I_n)$ all independent.
We then define a sample $Y_i = \sqrt{\rho} W_i \ones + \sqrt{1-\rho} Z_i$
so that the features have variance $1$ and equi-correlation $\rho$.
Finally, we construct our design matrix $X$ 
with a third-degree polynomial expansion of each $Y_{i,g}$ 
so that $X_{i,3g-2} = Y_{i,g}$, $X_{i,3g-1} = Y_{i,g}^2$, and $X_{i,3g} = Y_{i,g}^3$
for all $i\in [n]$ and $g \in [G]$,
so that $X \in \R^{n \times 3G}$.
Naturally, we define every 3 columns of $X$ as a group.
The natural parameter $\eta$ is generated from a linear model
$\eta = X \beta + \sigma \varepsilon$.
The first 6 components of $\beta$
are independently generated from a standard normal and the rest are set to 0.
The noise $\varepsilon \sim \Normal(0, I_n)$ and $\sigma$ is such that the signal-to-noise ratio is 3.
For the Gaussian loss, our response $y$ is identical to $\eta$,
and for the Binomial loss, we generate $y$ as Bernoulli random variables
where the logit parameters are given by $\eta$.
Finally, prior to fitting, we center and standardize the columns of $X$ and $y$.

In \Cref{fig:benchmark:group-lasso-sim},
we show the timings for solving the group lasso under 
the Gaussian loss (\Cref{fig:benchmark:group-lasso-sim:gaussian})
and the Binomial loss (\Cref{fig:benchmark:group-lasso-sim:binomial}).
We study a small ($n=100$) and large ($n=1000$) sample size case,
and for each case we vary the 
(equi-)correlation of the features ($\rho$).
Each plot shows the average times across $3$ trials
to solve the same path of 100 regularization values evenly-spaced on the log-scale 
starting at $\lambda_{\max}$,
the smallest $\lambda$ for which $\zeros$ is an optimal solution,
until $0.01 \lambda_{\max}$.
In all cases except one ($n=1000$, $\rho=0.95$, $G=5000$, Binomial loss), \texttt{adelie}
runs uniformly faster than the other packages.
\texttt{adelie} is up to 10 times faster than the next fastest package
and, more importantly, this difference factor remains to be the case 
as the data scales both in $n$ and $G$.
In the one case where \texttt{adelie} was not the fastest,
we observed that the screen set via the strong rule
pulled in a large number of groups from the very beginning.
This resulted in a large number of coordinate descent iterations
and therefore a relatively long runtime.
We are currently working to improve the strong rule to ameliorate such cases.

We briefly mention that we tried the following setting
with a larger group size of $s = 100$ rather than 3.
In this setting, we generate $Y \in \R^{n \times sG}$ as before
such that every feature is Gaussian with mean 0, variance 1, and equi-correlated with $\rho$.
We then group every $s$ features so that we have $G$ number of groups.
The response vector $y$ is generated in the same manner as before.
Every package except \texttt{adelie} struggled to converge
as their objective values were unacceptably far from that of \texttt{adelie};
we verified from the KKT conditions that \texttt{adelie} converged properly.
For this reason, we omit the timing comparison for this setting.

For the real datasets, we first expand
each feature (column) $x$ into three polynomial terms $x$, $x^2$, and $x^3$
just as with the Leukemia dataset in \Cref{sec:algorithm}
so that our feature matrix $X \in \R^{n \times 3G}$.
Prior to fitting, we center and standardize the columns of $X$ and $y$.
In \Cref{fig:benchmark:group-lasso-real},
we show the average timings to solve group lasso over 3 trials.
Similar to the simulation analysis,
we solve for the same path of 100 regularization values evenly-spaced on the log-scale for all packages.
Once again, we observe a dramatic speed improvement with \texttt{adelie}.

\subsection{Lasso Comparisons}\label{ssec:benchmark:lasso}

\begin{figure}[t]
    \centering 
    \begin{subfigure}{0.95\textwidth}
        \includegraphics[width=\linewidth]{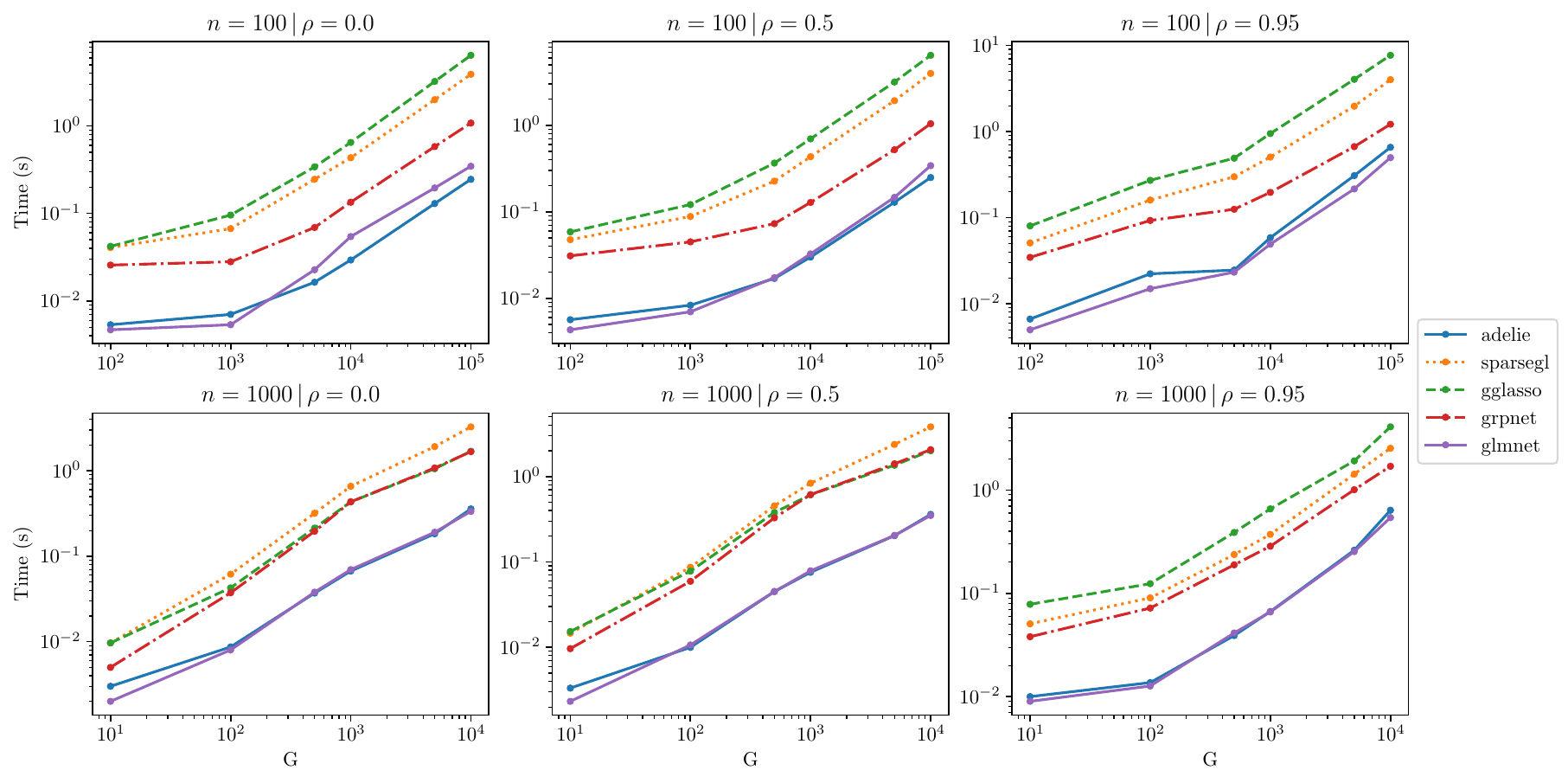}
        \caption{Gaussian Loss}
        \label{fig:benchmark:lasso-sim:gaussian}
    \end{subfigure}

    \begin{subfigure}{0.95\textwidth}
        \includegraphics[width=\linewidth]{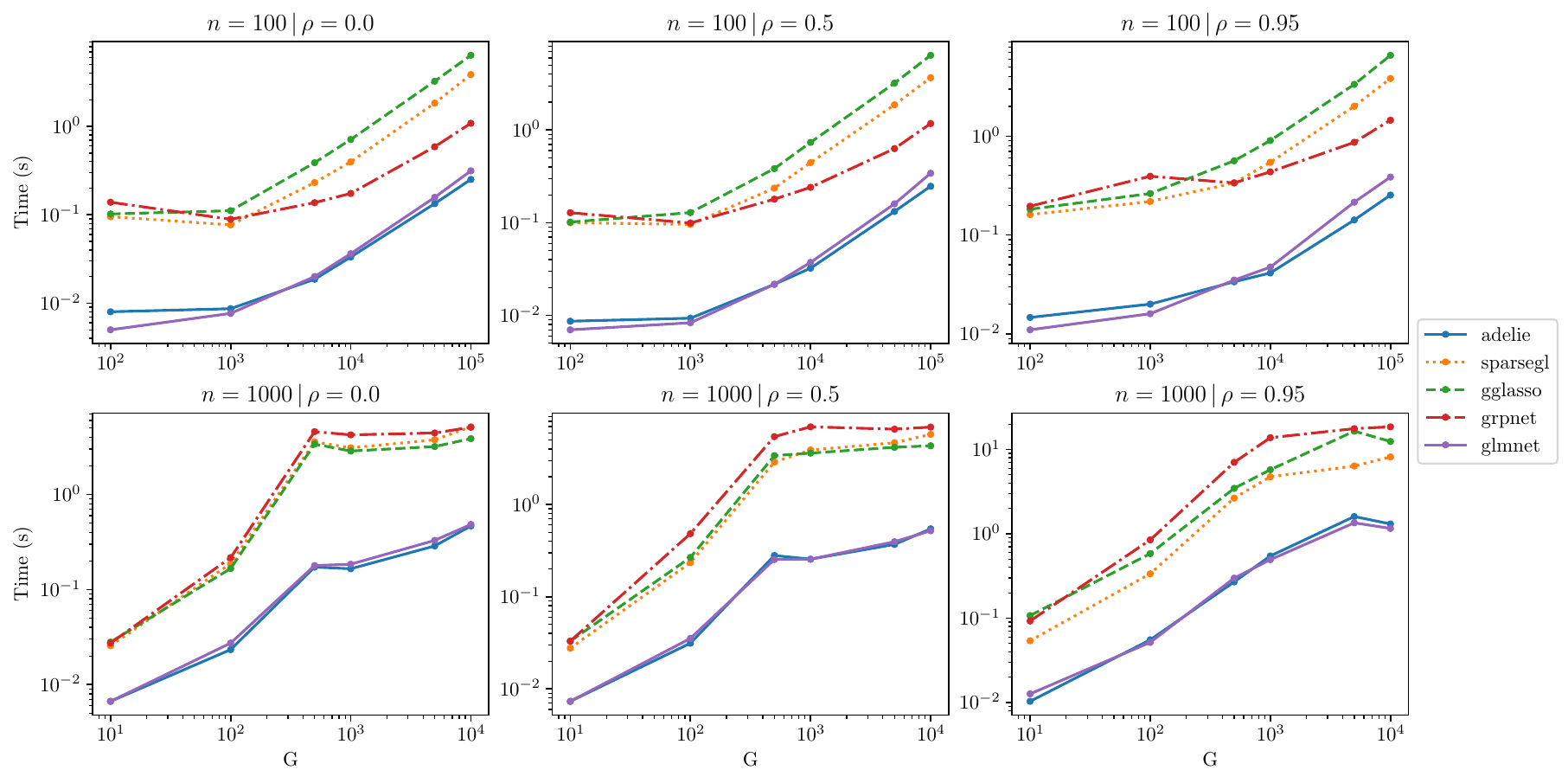}
        \caption{Binomial Loss}
        \label{fig:benchmark:lasso-sim:binomial}
    \end{subfigure}

    \caption{%
    Timing comparisons for solving the lasso under 
    the Gaussian loss (\Cref{fig:benchmark:lasso-sim:gaussian})
    and the Binomial loss (\Cref{fig:benchmark:lasso-sim:binomial})
    against existing R packages.
    We study a small ($n=100$) and large ($n=1000$) sample size case,
    and for each case we vary the 
    (equi-)correlation of the features ($\rho$).
    }
    \label{fig:benchmark:lasso-sim}
\end{figure}

\begin{figure}[t]
    \centering 
    \includegraphics[width=\linewidth]{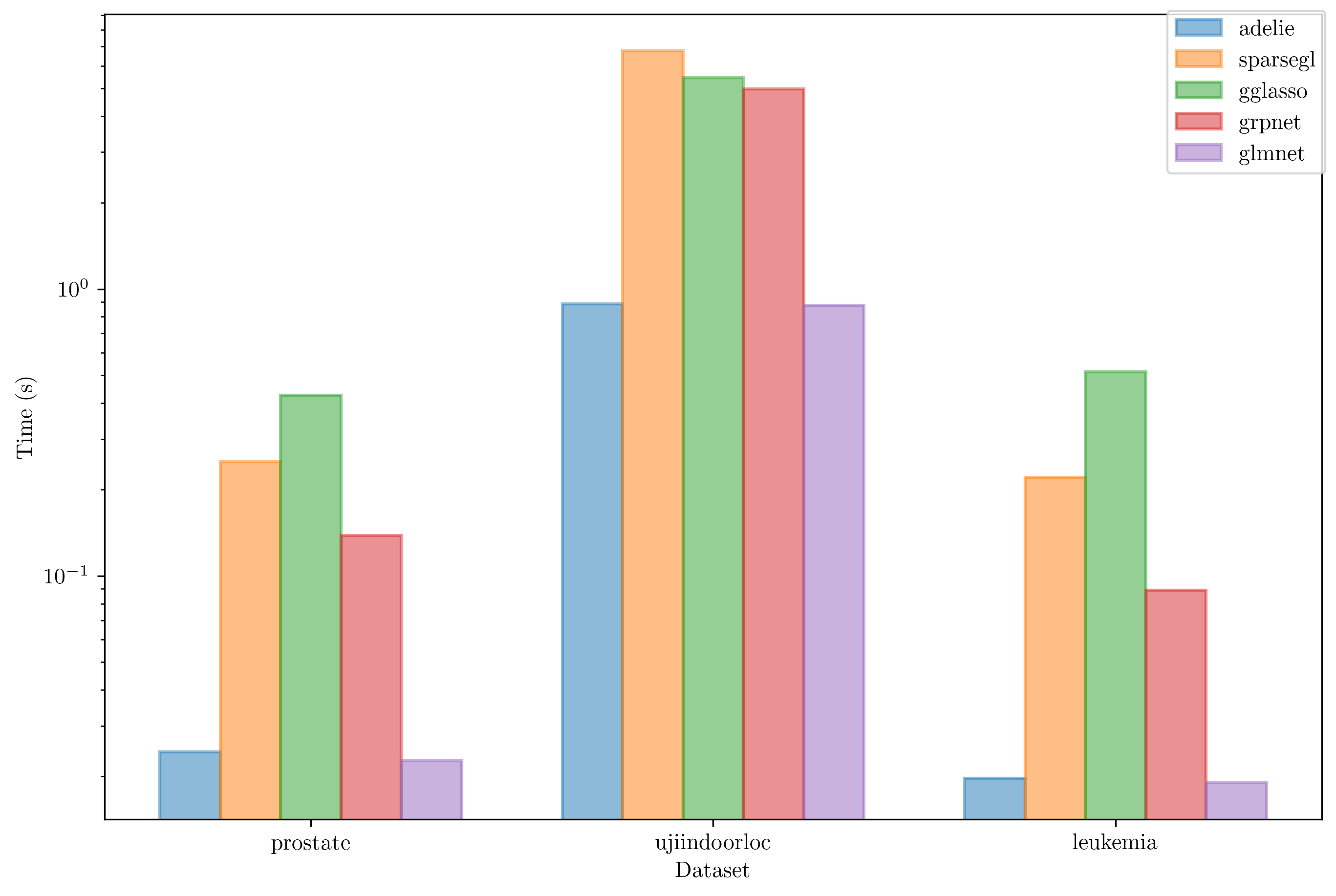}
    \caption{%
    Timing comparisons for solving the lasso 
    for the real datasets against existing R packages.
    }
    \label{fig:benchmark:lasso-real}
\end{figure}

In this section, we demonstrate that \texttt{adelie}
is a competitive lasso solver.
We also include \texttt{glmnet} \citep{friedman:2010}
in the list of packages for comparison.
As in \Cref{ssec:benchmark:group-lasso},
we show a timings comparison of solving the lasso
under the Gaussian and Binomial losses using 
simulated and real data.

Our simulation setup is taken from \citet{friedman:2010}.
For completion, we describe it next.
We begin by generating $n$ independent samples with $p$ features
from a Gaussian distribution
where every feature is equi-correlated with $\rho$.
The natural parameter $\eta = X \beta + \sigma \epsilon$
where $\beta_j = (-1)^j \exp(-2(j-1)/20)$,
$\epsilon \sim \Normal(0, 1)$,
and $\sigma$ is chosen such that the signal-to-noise ratio is $3$.
Note that the coefficients have alternating signs and decrease exponentially in magnitude.
For the Gaussian loss, we let our response $y \equiv \eta$
and for the Binomial loss, we generate $y$ from the Bernoulli distribution
where the logit parameter is given by $\eta$.
Prior to fitting, we center and standardize the columns of $X$ and $y$.

For the real datasets, we do not perform any special modifications 
to the data as in the group lasso setting
except for centering and standardizing the columns of $X$ and $y$.

In \Cref{fig:benchmark:lasso-sim,fig:benchmark:lasso-real},
we show the timings for solving the lasso under 
the Gaussian loss (\Cref{fig:benchmark:lasso-sim:gaussian})
and the Binomial loss (\Cref{fig:benchmark:lasso-sim:binomial})
for the simulated and real data setting, respectively.
We interpret the figures in the same way as in 
\Cref{fig:benchmark:group-lasso-sim,fig:benchmark:group-lasso-real}
for the group lasso comparisons in \Cref{ssec:benchmark:group-lasso}.
We note that \texttt{adelie} and \texttt{glmnet} have nearly identical timings 
and they clearly outperform the other packages.
This shows that \texttt{adelie} is not only a fast group lasso solver,
but also a fast lasso solver, suffering no loss of performance
compared to the highly optimized and specialized lasso solver \texttt{glmnet}.
\section{Discussion}\label{sec:discussion}

We demonstrate that block-coordinate descent is a powerful and effective method
for solving the group elastic net problem.
We derive a novel algorithm to solve the block-coordinate update for the Gaussian loss,
outperforming existing popular proximal gradient methods
in both performance and accuracy.
Moreover, our method naturally generalizes to other convex losses
via the proximal (quasi)-Newton method.
Through simulated and real data examples,
we demonstrate that our method is remarkably fast
for both group lasso and lasso problems.

Our Python package \texttt{adelie} is available under the MIT license
from PyPI \\
(\href{https://pypi.org/project/adelie}{https://pypi.org/project/adelie})
and GitHub (\href{https://github.com/JamesYang007/adelie}{https://github.com/JamesYang007/adelie}).
Our documentation is hosted on GitHub Pages
(\href{https://jamesyang007.github.io/adelie}{https://jamesyang007.github.io/adelie}).
The user can install the latest version by running 
\texttt{pip install adelie} in the terminal.
\section{Acknowledgments}\label{sec:acknowledgments}

Trevor Hastie was partially supported by grants DMS-2013736 from the National Science Foundation, 
and grant R01GM134483 from the National Institutes of Health. The authors thank Jonathan Taylor for
sharing his notes on the Cox model and other helpful comments, and Balasubramanian Narasimhan 
and Rob Tibshirani for helpful comments.
We also thank Kevin Fry, Kevin Guo, and Anav Sood for giving feedback on earlier drafts.

\bibliography{references}

\begin{thebibliography}{46}
\providecommand{\natexlab}[1]{#1}
\providecommand{\url}[1]{\texttt{#1}}
\expandafter\ifx\csname urlstyle\endcsname\relax
  \providecommand{\doi}[1]{doi: #1}\else
  \providecommand{\doi}{doi: \begingroup \urlstyle{rm}\Url}\fi

\bibitem[Beck and Teboulle(2009)]{beck:2009}
Beck, Amir and Teboulle, Marc.
\newblock A fast iterative shrinkage-thresholding algorithm for linear inverse
  problems.
\newblock \emph{SIAM Journal on Imaging Sciences}, 2\penalty0 (1):\penalty0
  183--202, 2009.
\newblock \doi{10.1137/080716542}.
\newblock URL \url{https://doi.org/10.1137/080716542}.

\bibitem[Boyd and Vandenberghe(2004)]{boyd:2004}
Boyd, Stephen and Vandenberghe, Lieven.
\newblock \emph{Convex optimization}.
\newblock Cambridge university press, 2004.

\bibitem[Breheny and Huang(2015)]{Breheny2015-zj}
Breheny, Patrick and Huang, Jian.
\newblock Group descent algorithms for nonconvex penalized linear and logistic
  regression models with grouped predictors.
\newblock \emph{Stat. Comput.}, 25\penalty0 (2):\penalty0 173--187, March 2015.

\bibitem[Brent(2013)]{brent:2013}
Brent, Richard~P.
\newblock \emph{Algorithms for minimization without derivatives}.
\newblock Courier Corporation, 2013.

\bibitem[Dekker(1969)]{dekker:1969}
Dekker, Theodorus~Jozef.
\newblock Finding a zero by means of successive linear interpolation.
\newblock \emph{Constructive aspects of the fundamental theorem of algebra},
  pages 37--51, 1969.

\bibitem[Friedman et~al.(2010)Friedman, Hastie, and Tibshirani]{friedman:2010}
Friedman, Jerome, Hastie, Trevor, and Tibshirani, Robert.
\newblock Regularization paths for generalized linear models via coordinate
  descent.
\newblock \emph{Journal of Statistical Software}, 33\penalty0 (1), 2010.
\newblock ISSN 1548-7660.
\newblock \doi{10.18637/jss.v033.i01}.
\newblock URL \url{http://dx.doi.org/10.18637/jss.v033.i01}.

\bibitem[Ghaoui et~al.(2011)Ghaoui, Viallon, and Rabbani]{ghaoui2011safe}
Ghaoui, Laurent~El, Viallon, Vivian, and Rabbani, Tarek.
\newblock Safe feature elimination for the lasso and sparse supervised learning
  problems, 2011.

\bibitem[Golub et~al.(1999)Golub, Slonim, Tamayo, Huard, Gaasenbeek, Mesirov,
  Coller, Loh, Downing, Caligiuri, Bloomfield, and Lander]{Golub1999}
Golub, T.~R., Slonim, D.~K., Tamayo, P., Huard, C., Gaasenbeek, M., Mesirov,
  J.~P., Coller, H., Loh, M.~L., Downing, J.~R., Caligiuri, M.~A., Bloomfield,
  C.~D., and Lander, E.~S.
\newblock Molecular classification of cancer: Class discovery and class
  prediction by gene expression monitoring.
\newblock \emph{Science}, 286\penalty0 (5439):\penalty0 531--537, October 1999.
\newblock ISSN 1095-9203.
\newblock \doi{10.1126/science.286.5439.531}.
\newblock URL \url{http://dx.doi.org/10.1126/science.286.5439.531}.

\bibitem[Hastie et~al.(2016)Hastie, Tibshirani, and Wainwright]{sls:2016}
Hastie, Trevor, Tibshirani, Rob, and Wainwright, Martin.
\newblock \emph{Statistical Learning with Sparsity}.
\newblock Chapman \& Hall, 2016.

\bibitem[He et~al.(2021)He, Le~Guen, Liu, Lee, Ma, Yang, Liu, Rutledge, Losada,
  Song, Belloy, Butler, Longo, Tang, Mormino, Wyss-Coray, Greicius, and
  Ionita-Laza]{He2021}
He, Zihuai, Le~Guen, Yann, Liu, Linxi, Lee, Justin, Ma, Shiyang, Yang,
  Andrew~C., Liu, Xiaoxia, Rutledge, Jarod, Losada, Patricia~Moran, Song,
  Bowen, Belloy, Michael~E., Butler, Robert~R., Longo, Frank~M., Tang, Hua,
  Mormino, Elizabeth~C., Wyss-Coray, Tony, Greicius, Michael~D., and
  Ionita-Laza, Iuliana.
\newblock Genome-wide analysis of common and rare variants via multiple
  knockoffs at biobank scale, with an application to alzheimer disease
  genetics.
\newblock \emph{The American Journal of Human Genetics}, 108\penalty0
  (12):\penalty0 2336--2353, December 2021.
\newblock ISSN 0002-9297.
\newblock \doi{10.1016/j.ajhg.2021.10.009}.
\newblock URL \url{http://dx.doi.org/10.1016/j.ajhg.2021.10.009}.

\bibitem[He et~al.(2022)He, Liu, Belloy, Le~Guen, Sossin, Liu, Qi, Ma, Gyawali,
  Wyss-Coray, Tang, Sabatti, Candès, Greicius, and Ionita-Laza]{He2022}
He, Zihuai, Liu, Linxi, Belloy, Michael~E., Le~Guen, Yann, Sossin, Aaron, Liu,
  Xiaoxia, Qi, Xinran, Ma, Shiyang, Gyawali, Prashnna~K., Wyss-Coray, Tony,
  Tang, Hua, Sabatti, Chiara, Candès, Emmanuel, Greicius, Michael~D., and
  Ionita-Laza, Iuliana.
\newblock Ghostknockoff inference empowers identification of putative causal
  variants in genome-wide association studies.
\newblock \emph{Nature Communications}, 13\penalty0 (1), November 2022.
\newblock ISSN 2041-1723.
\newblock \doi{10.1038/s41467-022-34932-z}.
\newblock URL \url{http://dx.doi.org/10.1038/s41467-022-34932-z}.

\bibitem[He et~al.(2024)He, Chu, Yang, Gu, Chen, Liu, Morrison, Belloy, Qi,
  Hejazi, Mathur, Le~Guen, Tang, Hastie, Ionita-laza, Sabatti, and
  Candès]{He2024}
He, Zihuai, Chu, Benjamin, Yang, James, Gu, Jiaqi, Chen, Zhaomeng, Liu, Linxi,
  Morrison, Tim, Belloy, Michael~E., Qi, Xinran, Hejazi, Nima, Mathur, Maya,
  Le~Guen, Yann, Tang, Hua, Hastie, Trevor, Ionita-laza, Iuliana, Sabatti,
  Chiara, and Candès, Emmanuel.
\newblock In silico identification of putative causal genetic variants.
\newblock March 2024.
\newblock \doi{10.1101/2024.02.28.582621}.
\newblock URL \url{http://dx.doi.org/10.1101/2024.02.28.582621}.

\bibitem[Helwig(2024)]{Helwig:2024}
Helwig, Nathaniel~E.
\newblock \emph{grpnet: Group Elastic Net Regularized GLMs and GAMs}, 2024.
\newblock URL \url{https://CRAN.R-project.org/package=grpnet}.
\newblock R package version 0.3.

\bibitem[{Joaqun Torres-Sospedra, Raul Montoliu, Adolfo Martnez-Us, Tomar
  Arnau, Joan Avariento}(2014)]{ujiindoorloc}
{Joaqun Torres-Sospedra, Raul Montoliu, Adolfo Martnez-Us, Tomar Arnau, Joan
  Avariento}.
\newblock {UJIIndoorLoc}, 2014.

\bibitem[Klosa et~al.(2020)Klosa, Simon, Westermark, Liebscher, and
  Wittenburg]{klosa:2020}
Klosa, J., Simon, N., Westermark, P.O., Liebscher, V., and Wittenburg, D.
\newblock Seagull: lasso, group lasso and sparse-group lasso regularization for
  linear regression models via proximal gradient descent.
\newblock \emph{BMC Bioinformatics}, 21\penalty0 (407), 2020.
\newblock \doi{10.1186/s12859-020-03725-w}.

\bibitem[Krishnapuram et~al.(2005)Krishnapuram, Carin, Figueiredo, and
  Hartemink]{Krishnapuram2005}
Krishnapuram, B., Carin, L., Figueiredo, M.A.T., and Hartemink, A.J.
\newblock Sparse multinomial logistic regression: fast algorithms and
  generalization bounds.
\newblock \emph{IEEE Transactions on Pattern Analysis and Machine
  Intelligence}, 27\penalty0 (6):\penalty0 957--968, June 2005.
\newblock ISSN 2160-9292.
\newblock \doi{10.1109/tpami.2005.127}.
\newblock URL \url{http://dx.doi.org/10.1109/TPAMI.2005.127}.

\bibitem[Lee et~al.(2014)Lee, Sun, and Saunders]{Lee2014}
Lee, Jason~D., Sun, Yuekai, and Saunders, Michael~A.
\newblock Proximal newton-type methods for minimizing composite functions.
\newblock \emph{SIAM Journal on Optimization}, 24\penalty0 (3):\penalty0
  1420--1443, January 2014.
\newblock ISSN 1095-7189.
\newblock \doi{10.1137/130921428}.
\newblock URL \url{http://dx.doi.org/10.1137/130921428}.

\bibitem[Li et~al.(2020)Li, Chang, Justesen, Tanigawa, Qian, Hastie, Rivas, and
  Tibshirani]{Li2020}
Li, Ruilin, Chang, Christopher, Justesen, Johanne~M, Tanigawa, Yosuke, Qian,
  Junyang, Hastie, Trevor, Rivas, Manuel~A, and Tibshirani, Robert.
\newblock Fast lasso method for large-scale and ultrahigh-dimensional cox model
  with applications to uk biobank.
\newblock \emph{Biostatistics}, 23\penalty0 (2):\penalty0 522--540, September
  2020.
\newblock ISSN 1468-4357.
\newblock \doi{10.1093/biostatistics/kxaa038}.
\newblock URL \url{http://dx.doi.org/10.1093/biostatistics/kxaa038}.

\bibitem[Li et~al.(2021)Li, Chang, Tanigawa, Narasimhan, Hastie, Tibshirani,
  and Rivas]{Li2021}
Li, Ruilin, Chang, Christopher, Tanigawa, Yosuke, Narasimhan, Balasubramanian,
  Hastie, Trevor, Tibshirani, Robert, and Rivas, Manuel~A.
\newblock Fast numerical optimization for genome sequencing data in population
  biobanks.
\newblock \emph{Bioinformatics}, 37\penalty0 (22):\penalty0 4148--4155, June
  2021.
\newblock ISSN 1367-4811.
\newblock \doi{10.1093/bioinformatics/btab452}.
\newblock URL \url{http://dx.doi.org/10.1093/bioinformatics/btab452}.

\bibitem[Liang et~al.(2022)Liang, Cohen, Heinsfeld, Pestilli, and
  McDonald]{sparsegl:2022}
Liang, Xiaoxuan, Cohen, Aaron, Heinsfeld, Anibal~Solón, Pestilli, Franco, and
  McDonald, Daniel~J.
\newblock sparsegl: An r package for estimating sparse group lasso, 2022.
\newblock URL \url{https://arxiv.org/abs/2208.02942}.

\bibitem[Loris(2009)]{loris:2009}
Loris, Ignace.
\newblock On the performance of algorithms for the minimization of
  $\ell_1$-penalized functionals.
\newblock \emph{Inverse Problems}, 25\penalty0 (3):\penalty0 035008, jan 2009.
\newblock \doi{10.1088/0266-5611/25/3/035008}.
\newblock URL \url{https://dx.doi.org/10.1088/0266-5611/25/3/035008}.

\bibitem[McCullagh and Nelder(1989)]{McCullagh1989}
McCullagh, P. and Nelder, J.~A.
\newblock \emph{Generalized Linear Models}.
\newblock Springer US, 1989.
\newblock ISBN 9781489932426.
\newblock \doi{10.1007/978-1-4899-3242-6}.
\newblock URL \url{http://dx.doi.org/10.1007/978-1-4899-3242-6}.

\bibitem[McDonald et~al.(2023)McDonald, Liang, {Solón Heinsfeld}, and
  Cohen]{sparsegl:R:2023}
McDonald, Daniel~J., Liang, Xiaoxuan, {Solón Heinsfeld}, Anibal, and Cohen,
  Aaron.
\newblock \emph{sparsegl: Sparse Group Lasso}, 2023.
\newblock URL \url{https://CRAN.R-project.org/package=sparsegl}.
\newblock R package version 1.0.2.

\bibitem[Meier(2020)]{grplasso:R:2020}
Meier, Lukas.
\newblock \emph{grplasso: Fitting User-Specified Models with Group Lasso
  Penalty}, 2020.
\newblock URL \url{https://CRAN.R-project.org/package=grplasso}.
\newblock R package version 0.4-7.

\bibitem[Meier et~al.(2008)Meier, Van De~Geer, and B\"{o}hlmann]{meier:2008}
Meier, Lukas, Van De~Geer, Sara, and B\"{o}hlmann, Peter.
\newblock The group lasso for logistic regression.
\newblock \emph{Journal of the Royal Statistical Society B}, 70\penalty0
  (1):\penalty0 53--71, 2008.
\newblock \doi{10.1111/j.1467-9868.2007.00627.x}.

\bibitem[Mills and Rahal(2019)]{Mills2019}
Mills, Melinda~C. and Rahal, Charles.
\newblock A scientometric review of genome-wide association studies.
\newblock \emph{Communications Biology}, 2\penalty0 (1), January 2019.
\newblock ISSN 2399-3642.
\newblock \doi{10.1038/s42003-018-0261-x}.
\newblock URL \url{http://dx.doi.org/10.1038/s42003-018-0261-x}.

\bibitem[O'Donoghue and Cand\'{e}s(2015)]{odonoghue:2015}
O'Donoghue, Brendan and Cand\'{e}s, Emmanuel.
\newblock Adaptive restart for accelerated gradient schemes.
\newblock \emph{Foundations of Computational Mathematics}, 15:\penalty0
  715--732, 2015.
\newblock
  \doi{https://doi-org.stanford.idm.oclc.org/10.1007/s10208-013-9150-3}.

\bibitem[Qian et~al.(2020)Qian, Tanigawa, Du, Aguirre, Chang, Tibshirani,
  Rivas, and Hastie]{Qian2020}
Qian, Junyang, Tanigawa, Yosuke, Du, Wenfei, Aguirre, Matthew, Chang, Chris,
  Tibshirani, Robert, Rivas, Manuel~A., and Hastie, Trevor.
\newblock A fast and scalable framework for large-scale and
  ultrahigh-dimensional sparse regression with application to the uk biobank.
\newblock \emph{PLOS Genetics}, 16\penalty0 (10):\penalty0 e1009141, October
  2020.
\newblock ISSN 1553-7404.
\newblock \doi{10.1371/journal.pgen.1009141}.
\newblock URL \url{http://dx.doi.org/10.1371/journal.pgen.1009141}.

\bibitem[Simon and Tibshirani(2012)]{simon:2012}
Simon, Noah and Tibshirani, Robert.
\newblock Standardization and the group lasso penalty.
\newblock \emph{Statistica Sinica}, 22\penalty0 (3):\penalty0 983--1001, 2012.
\newblock ISSN 10170405, 19968507.
\newblock URL \url{http://www.jstor.org/stable/24309971}.

\bibitem[Simon et~al.(2013{\natexlab{a}})Simon, Friedman, Hastie, and
  Tibshirani]{Simon2013-yv}
Simon, Noah, Friedman, Jerome, Hastie, Trevor, and Tibshirani, Robert.
\newblock A {Sparse-Group} lasso.
\newblock \emph{J. Comput. Graph. Stat.}, 22\penalty0 (2):\penalty0 231--245,
  April 2013{\natexlab{a}}.

\bibitem[Simon et~al.(2019)Simon, Friedman, Hastie, and Tibshirani]{SGL:R:2019}
Simon, Noah, Friedman, Jerome, Hastie, Trevor, and Tibshirani, Rob.
\newblock \emph{SGL: Fit a GLM (or Cox Model) with a Combination of Lasso and
  Group Lasso Regularization}, 2019.
\newblock URL \url{https://CRAN.R-project.org/package=SGL}.
\newblock R package version 1.3.

\bibitem[Simon et~al.(2013{\natexlab{b}})Simon, Friedman, and
  Hastie]{Simon2013ABD}
Simon, Noah~R., Friedman, Jerome~H., and Hastie, Trevor~J.
\newblock A blockwise descent algorithm for group-penalized multiresponse and
  multinomial regression.
\newblock \emph{arXiv: Computation}, 2013{\natexlab{b}}.
\newblock URL \url{https://api.semanticscholar.org/CorpusID:18386302}.

\bibitem[Singh et~al.(2002)Singh, Febbo, Ross, Jackson, Manola, Ladd, Tamayo,
  Renshaw, D'Amico, Richie, Lander, Loda, Kantoff, Golub, and
  Sellers]{Singh2002-gv}
Singh, Dinesh, Febbo, Phillip~G, Ross, Kenneth, Jackson, Donald~G, Manola,
  Judith, Ladd, Christine, Tamayo, Pablo, Renshaw, Andrew~A, D'Amico,
  Anthony~V, Richie, Jerome~P, Lander, Eric~S, Loda, Massimo, Kantoff,
  Philip~W, Golub, Todd~R, and Sellers, William~R.
\newblock Gene expression correlates of clinical prostate cancer behavior.
\newblock \emph{Cancer Cell}, 1\penalty0 (2):\penalty0 203--209, March 2002.

\bibitem[Sudlow et~al.(2015)Sudlow, Gallacher, Allen, Beral, Burton, Danesh,
  Downey, Elliott, Green, Landray, Liu, Matthews, Ong, Pell, Silman, Young,
  Sprosen, Peakman, and Collins]{Sudlow2015}
Sudlow, Cathie, Gallacher, John, Allen, Naomi, Beral, Valerie, Burton, Paul,
  Danesh, John, Downey, Paul, Elliott, Paul, Green, Jane, Landray, Martin, Liu,
  Bette, Matthews, Paul, Ong, Giok, Pell, Jill, Silman, Alan, Young, Alan,
  Sprosen, Tim, Peakman, Tim, and Collins, Rory.
\newblock Uk biobank: An open access resource for identifying the causes of a
  wide range of complex diseases of middle and old age.
\newblock \emph{PLOS Medicine}, 12\penalty0 (3):\penalty0 e1001779, March 2015.
\newblock ISSN 1549-1676.
\newblock \doi{10.1371/journal.pmed.1001779}.
\newblock URL \url{http://dx.doi.org/10.1371/journal.pmed.1001779}.

\bibitem[Tibshirani(1996)]{Tibshirani1996}
Tibshirani, Robert.
\newblock Regression shrinkage and selection via the lasso.
\newblock \emph{Journal of the Royal Statistical Society. Series B
  (Methodological)}, 58\penalty0 (1):\penalty0 267--288, 1996.
\newblock ISSN 00359246.
\newblock URL \url{http://www.jstor.org/stable/2346178}.

\bibitem[Tibshirani et~al.(2012)Tibshirani, Bien, Friedman, Hastie, Simon,
  Taylor, and Tibshirani]{Tibshirani2012Strong}
Tibshirani, Robert, Bien, Jacob, Friedman, Jerome, Hastie, Trevor, Simon, Noah,
  Taylor, Jonathan, and Tibshirani, Ryan~J.
\newblock Strong rules for discarding predictors in lasso-type problems.
\newblock \emph{Journal of the Royal Statistical Society. Series B (Statistical
  Methodology)}, 74\penalty0 (2):\penalty0 245--266, 2012.
\newblock ISSN 13697412, 14679868.
\newblock URL \url{http://www.jstor.org/stable/41430939}.

\bibitem[Tseng(2001)]{tseng:2001}
Tseng, P.
\newblock Convergence of a block coordinate descent method for
  nondifferentiable minimization.
\newblock \emph{Journal of Optimization Theory and Applications}, 109:\penalty0
  475--494, 2001.
\newblock \doi{https://doi-org.stanford.idm.oclc.org/10.1023/A:1017501703105}.

\bibitem[Uffelmann et~al.(2021)Uffelmann, Huang, Munung, de~Vries, Okada,
  Martin, Martin, Lappalainen, and Posthuma]{Uffelmann2021}
Uffelmann, Emil, Huang, Qin~Qin, Munung, Nchangwi~Syntia, de~Vries, Jantina,
  Okada, Yukinori, Martin, Alicia~R., Martin, Hilary~C., Lappalainen, Tuuli,
  and Posthuma, Danielle.
\newblock Genome-wide association studies.
\newblock \emph{Nature Reviews Methods Primers}, 1\penalty0 (1), August 2021.
\newblock ISSN 2662-8449.
\newblock \doi{10.1038/s43586-021-00056-9}.
\newblock URL \url{http://dx.doi.org/10.1038/s43586-021-00056-9}.

\bibitem[Wang et~al.(2015)Wang, Wonka, and Ye]{wang:2015}
Wang, Jie, Wonka, Peter, and Ye, Jieping.
\newblock Lasso screening rules via dual polytope projection.
\newblock \emph{Journal of Machine Learning Research}, 16\penalty0
  (32):\penalty0 1063--1101, 2015.
\newblock URL \url{http://jmlr.org/papers/v16/wang15a.html}.

\bibitem[Witte(2010)]{Witte2010}
Witte, John~S.
\newblock Genome-wide association studies and beyond.
\newblock \emph{Annual Review of Public Health}, 31\penalty0 (1):\penalty0
  9--20, March 2010.
\newblock ISSN 1545-2093.
\newblock \doi{10.1146/annurev.publhealth.012809.103723}.
\newblock URL \url{http://dx.doi.org/10.1146/annurev.publhealth.012809.103723}.

\bibitem[Wright et~al.(2009)Wright, Nowak, and Figueiredo]{wright:2009}
Wright, Stephen~J., Nowak, Robert~D., and Figueiredo, MÁrio A.~T.
\newblock Sparse reconstruction by separable approximation.
\newblock \emph{IEEE Transactions on Signal Processing}, 57\penalty0
  (7):\penalty0 2479--2493, 2009.
\newblock \doi{10.1109/TSP.2009.2016892}.

\bibitem[Yang and Zou(2015)]{Yang2015-zx}
Yang, Yi and Zou, Hui.
\newblock A fast unified algorithm for solving group-lasso penalize learning
  problems.
\newblock \emph{Stat. Comput.}, 25\penalty0 (6):\penalty0 1129--1141, November
  2015.

\bibitem[Yang et~al.(2024)Yang, Zou, and Bhatnagar]{gglasso:R:2024}
Yang, Yi, Zou, Hui, and Bhatnagar, Sahir.
\newblock \emph{gglasso: Group Lasso Penalized Learning Using a Unified BMD
  Algorithm}, 2024.
\newblock URL \url{https://CRAN.R-project.org/package=gglasso}.
\newblock R package version 1.5.1.

\bibitem[Yuan and Lin(2006)]{yuan:2006}
Yuan, Ming and Lin, Yi.
\newblock Model selection and estimation in regression with grouped variables.
\newblock \emph{Journal of the Royal Statistical Society B}, 68\penalty0
  (1):\penalty0 49--67, 2006.
\newblock \doi{10.1111/j.1467-9868.2005.00532.x}.

\bibitem[Zeng and Breheny(2018)]{zeng2018biglasso}
Zeng, Yaohui and Breheny, Patrick.
\newblock The biglasso package: A memory- and computation-efficient solver for
  lasso model fitting with big data in r, 2018.

\bibitem[Zou and Hastie(2005)]{zou:2005}
Zou, Hui and Hastie, Trevor.
\newblock Regularization and variable selection via the elastic net.
\newblock \emph{Journal of the Royal Statistical Society. Series B (Statistical
  Methodology)}, 67\penalty0 (2):\penalty0 301--320, 2005.
\newblock ISSN 13697412, 14679868.
\newblock URL \url{http://www.jstor.org/stable/3647580}.

\end{thebibliography}

\appendix

\section{Proof of \Cref{thm:algorithm:existence}}%
\label{app:thm:algorithm:existence}

Consider the optimization problem \labelcref{eq:algorithm:bcd-general} with $\lambda > 0$.
Suppose that $v_S = \zeros$ where $S$ is given in \labelcref{eq:algorithm:bcd-general-ass1}.
Without loss of generality, we may assume $S^c$ is non-empty 
so that there exists at least one $i$ such that $\Sigma_{ii} > 0$.
Otherwise, by hypothesis, we must have $v = \zeros$ so that \labelcref{eq:algorithm:bcd-general}
reduces to minimizing $\norm{x}_2$, which is uniquely minimized at $x^\star = \zeros$.

By the KKT conditions, the (finite) minimizer $x^\star$ must satisfy
\begin{align*}
    \Sigma x - v + \lambda u = \zeros
\end{align*}
where $u$ is a sub-gradient of $\norm{\cdot}_2$ at $x$.
We first prove the identity \labelcref{eq:algorithm:block-solution}.
If $x^\star = \zeros$, then we must have that $\norm{v}_2 \leq \lambda$.
Otherwise, $u \equiv x / \norm{x}_2$ so that
\begin{align}
    x = \pr{\Sigma + \frac{\lambda}{\norm{x}_2} I}^{-1} v 
    \label{eq:app:block-solution}
\end{align}
It remains to show that if $\norm{v}_2 \leq \lambda$, then $x^\star = \zeros$,
which will establish the identity \labelcref{eq:algorithm:block-solution}.

In that endeavor, suppose $x^\star \neq \zeros$.
Then, taking norms on both sides of \labelcref{eq:app:block-solution}, 
we have that $x^\star$ satisfies
\begin{align*}
    \sum\limits_{i=1}^d 
    \frac{v_i^2}{(\Sigma_{ii} \norm{x}_2 + \lambda)^2}
    =
    1
\end{align*}
Note that we then have that $\norm{v}_2^2 \lambda^{-2} \geq 1$,
or equivalently, $\norm{v}_2 \geq \lambda$.
There must exist some $i$ such that $v_i \neq 0$ and $\Sigma_{ii} > 0$.
Otherwise, $v \equiv \zeros$ by hypothesis that $v_S = \zeros$,
which contradicts $\norm{v}_2 \geq \lambda > 0$.
Consequently, for the element $i$ such that $v_i \neq 0$ and $\Sigma_{ii} > 0$, 
we have the strict inequality
\begin{align*}
    \frac{v_i^2}{(\Sigma_{ii} \norm{x}_2 + \lambda)^2}
    <
    v_i^2 \lambda^{-2}
\end{align*}
Hence, we have that $\norm{v}_2 > \lambda$.
This establishes the equivalence $x^\star = \zeros \iff \norm{v}_2 \leq \lambda$.

\section{Proof of \Cref{thm:algorithm:newton:newton-convex}}%
\label{app:thm:algorithm:newton:newton-convex}

Consider any $f$ satisfying the hypothesis of \Cref{thm:algorithm:newton:newton-convex}.
Let $x^{(0)}$ be any initial point such that $f(x^{(0)}) \geq 0$.
For a given $k\geq 0$, suppose $f(x^{(k)}) \geq 0$.
Then, by convexity of $f$ and the definition of $x^{(k+1)}$,
\begin{align*}
    f(x^{(k+1)})
    &\geq
    f(x^{(k)})
    + f'(x^{(k)}) (x^{(k+1)} - x^{(k)})
    =
    0
\end{align*}
By induction, we have that $f(x^{(k)}) \geq 0$ for every $k \geq 0$.
Since $f$ is decreasing and has a non-vanishing derivative, 
we must have that $f'(x) < 0$ for all $x$.
In particular, $x^{(k+1)} \geq x^{(k)}$ so that it is an increasing sequence.
Note that if $f(x^{(k)}) = 0$ for some $k$, then $x^{(k+j)} = x^{(k)}$ for all $j \geq 0$
so that the sequence converges in a trivial fashion to a root of $f$.
Hence, without loss of generality, suppose $f(x^{(k)}) > 0$ for all $k \geq 0$.
By the decreasing property of $f$, 
we must then have that $x^{(k)} \leq r$ for every $k\geq 0$.
Since the sequence is increasing and bounded above,
it converges to some value $x^{(\infty)}$.
By continuity, taking $k\to\infty$ in \labelcref{eq:thm:algorithm:newton:newton},
\begin{align*}
    x^{(\infty)} = x^{(\infty)} - \frac{f(x^{(\infty)})}{f'(x^{(\infty)})}
    \implies
    f(x^{(\infty)}) = 0
\end{align*}
which shows that $x^{(\infty)}$ is a root of $f$.

\section{Properties of $\varphi$ in \labelcref{eq:algorithm:newton:varphi-def}}
\label{app:algorithm:newton:varphi-def}

Consider $\varphi$ as in \labelcref{eq:algorithm:newton:varphi-def}.
Recall by \Cref{thm:algorithm:existence}, we only consider $\varphi$
when $\norm{v}_2 > \lambda$.
Similar to the proof of \Cref{thm:algorithm:existence} in \Cref{app:thm:algorithm:existence},
we have that there exists some $i$ such that $v_i \neq 0$ and $\Sigma_{ii} > 0$.
Then, differentiating $\varphi$, we get that
\begin{align*}
    \varphi'(h) 
    =
    -2\sum_{i=1}^d \frac{v_i^2 \Sigma_{ii}}{(\Sigma_{ii} h + \lambda)^3}
    < 0
    ,\quad
    \varphi''(h)
    =
    6\sum_{i=1}^d
    \frac{v_i^2 \Sigma_{ii}^2}{(\Sigma_{ii} h + \lambda)^4}
    > 0
\end{align*}
This shows that $\varphi$ is strictly decreasing and strictly convex.

Next, rewriting $\varphi$ as
\begin{align}
    \varphi(h)
    =
    \sum_{i \notin S} \frac{v_i^2}{(\Sigma_{ii} h + \lambda)^2}
    - 1
\end{align}
we see that $\lim_{h\to\infty} \varphi(h) < 0$.
Note that $\varphi(0) = \norm{v}_2^2 / \lambda^2 -1 > 0$.
Hence, a finite positive root uniquely exists for $\varphi$.

\section{Derivation of $h_\star$ and $h^\star$}
\label{app:algorithm:newton-abs:lower-upper}

We begin with the lower bound $h_\star$.
Let $a(h), b(h) \in \R^d$ be defined by $a_k(h) := \Sigma_{kk} h + \lambda$
and $b_k(h) := \abs{v_k}/a_k(h)$ for each $k=1,\ldots, d$.
Then, by Cauchy-Schwarz,
\begin{align*}
    \norm{v}_1
    &:=
    \sum\limits_{k=1}^d \abs{v_k}
    =
    \sum\limits_{k=1}^d a_k(h) b_k(h)
    \leq
    \norm{a(h)}_{2} \norm{b(h)}_2
\end{align*}
Hence, if $\norm{a(h)}_2 \leq \norm{v}_1$,
then $\norm{b(h)}_2 \geq 1$, or equivalently, $\varphi(h) \geq 0$.
We see that $\norm{a(h)}_2 \leq \norm{v}_1$ if and only if
\begin{align*}
    \sum\limits_{i=1}^d
    (\Sigma_{ii} h + \lambda)^2
    \leq
    \norm{v}_1^2
\end{align*}
By the quadratic formula, we may choose the largest $h$ that satisfies the above condition,
which is given by \labelcref{eq:algorithm:newton-abs:lower}.

The upper bound $h^\star$ is even easier.
Since
\begin{align*}
    \sum\limits_{i=1}^d
    \frac{v_i^2}{(\Sigma_{ii} h + \lambda)^2}
    &=
    \sum\limits_{i: \Sigma_{ii} > 0}
    \frac{v_i^2}{(\Sigma_{ii} h + \lambda)^2}
    \leq 
    h^{-2}
    \sum\limits_{i: \Sigma_{ii} > 0}
    \frac{v_i^2}{\Sigma_{ii}^2 }
\end{align*}
we may set $h^\star$ to \labelcref{eq:algorithm:newton-abs:upper}.
Then, we immediately have that $\varphi(h^\star) \leq 0$.

\end{document}